\documentclass[a4paper,fleqn]{cas-dc}

\usepackage{xparse}%
\usepackage{blindtext}
\usepackage{morewrites}
\usepackage{wrapfig}
\usepackage{xkeyval}
\usepackage[authoryear]{natbib}
\newcommand\ChangeRT[1]{\noalign{\hrule height #1}}
\usepackage{makecell}
\usepackage{amsmath,amssymb,amsfonts}
\usepackage{algorithmic}
\usepackage{graphicx}
\usepackage{float}
\usepackage{booktabs}
\usepackage{soul, color, xcolor}

\usepackage{listings}
\usepackage{picinpar}
\usepackage{bbding}
\usepackage{caption}

\usepackage{pifont}

\usepackage{amssymb}
\definecolor{lightblue_qing}{RGB}{47,165,212}
\usepackage{wrapfig}
\usepackage{hyperref}
\hypersetup{
    colorlinks=true,
    linkcolor=lightblue_qing,
    filecolor=magenta,      
    urlcolor=cyan,
    citecolor= lightblue_qing,
    pdftitle={Overleaf Example},
    pdfpagemode=FullScreen,
}

\newcommand{\setParDis}{\setlength {\parskip} {0.3cm} }
\newcommand{\setParDef}{\setlength {\parskip} {0pt} }

\usepackage{wasysym}

\usepackage{url}

\def\tsc#1{\csdef{#1}{\textsc{\lowercase{#1}}\xspace}}
\tsc{WGM}
\tsc{QE}
\tsc{EP}
\tsc{PMS}
\tsc{BEC}
\tsc{DE}

\begin{document}
\let\WriteBookmarks\relax
\def\floatpagepagefraction{1}
\def\textpagefraction{.001}
\let\printorcid\relax


\shorttitle{A Survey of DeFi Security: Challenges and Opportunities}

\shortauthors{Wenkai Li et~al.}

\title [mode = title]{\textbf{A Survey of DeFi Security: Challenges and Opportunities}}  





%

\author[1]{Wenkai Li}[type=editor,
                        style=chinese,
                        auid=000,
                        bioid=1]



\author[1]{Jiuyang Bu}[type=editor,
                        style=chinese,
                        auid=000,
                        bioid=1]

\author[1]{Xiaoqi Li}[type=editor,
                        style=chinese,
                        auid=000,
                        bioid=1] 
\cormark[1]



\author[1]{Hongli Peng}[type=editor,
                        style=chinese,
                        auid=000,
                        bioid=1]

\author[1]{Yuanzheng Niu}[type=editor,
                        style=chinese,
                        auid=000,
                        bioid=1]

\author[1,2]{Yuqing Zhang}[type=editor,
                        style=chinese,
                        auid=000,
                        bioid=1]

\address[1]{School of Cyberspace Security, Hainan University, Renmin Avenue 58, Haikou, 570228, China}
\address[2]{National Computer Network Intrusion Protection Center, University of Chinese Academy of Sciences, Yuquan Road 19, Beijing, 100049, China}

\cortext[cor1]{Corresponding author.}
\nonumnote{Email: liwenkai871@gmail.com (W. Li), csxqli@gmail.com (X. Li)}

\begin{abstract}
DeFi, or Decentralized Finance, is based on a distributed ledger called blockchain technology. Using blockchain, DeFi may customize the execution of predetermined operations between parties. The DeFi system use blockchain technology to execute user transactions, such as lending and exchanging. The total value locked in DeFi decreased from \$200 billion in April 2022 to \$80 billion in July 2022, indicating that security in this area remained problematic. In this paper, we address the deficiency in DeFi security studies. To our best knowledge, our paper is the first to make a systematic analysis of DeFi security.  First, we summarize the DeFi-related vulnerabilities in each blockchain layer. Additionally, application-level vulnerabilities are also analyzed. Then we classify and analyze real-world DeFi attacks based on the principles that correlate to the vulnerabilities. In addition, we collect optimization strategies from the data, network, consensus, smart contract, and application layers. And then, we describe the weaknesses and technical approaches they address. On the basis of this comprehensive analysis, we summarize several challenges and possible future directions in DeFi to offer ideas for further research.
\end{abstract}


\maketitle

\section{Introduction}
\label{sec:introduction}

The blockchain concept originated from the research of \cite{haber1990time} added timestamps to text, audio, and video files in digital form to guarantee their authenticity.
When \cite{nakamoto2008bitcoin} refined the blockchain concept for the first time, blockchain began to serve as a decentralized network with numerous properties, attracting considerable research.
At the same time, the application of cryptography principles~\citep{nakamoto2008bitcoin} and the promotion of consensus mechanisms~\citep{jakobsson1999proofs} have enabled digital currencies with blockchain as the core to allow untrusting parties to complete transactions securely.

Suppose blockchain-based Bitcoin transactions represent the blockchain 1.0 era. In that case, the combination of smart contracts and blockchain signifies the era of blockchain 2.0. \cite{szabo1996smart} first introduced the concept of the smart contract, which denoted a promise or agreement in digital form. \cite{buterin2014next} proposed Ethereum, which updates and verifies blockchain data via the state. Ethereum is currently a significant platform for smart contracts and decentralized applications. The emergence of the Ethereum platform has stimulated the emergence of various blockchain platforms, such as BNB and Polygon. 

In addition, Decentralized Finance (DeFi) is a decentralized application that uses blockchain in the financial domain to implement pre-defined financial protocols. Blockchain technology is widely used in various fields, such as education, health, and finance. Moreover, since the Ethereum blockchain technology integrated with finance more effectively during the Bitcoin era, DeFi technology in the financial field is gaining more attention.

\begin{figure}[htbp]
\setlength{\abovecaptionskip}{0cm} 
	\centering
		\includegraphics[width=8cm]{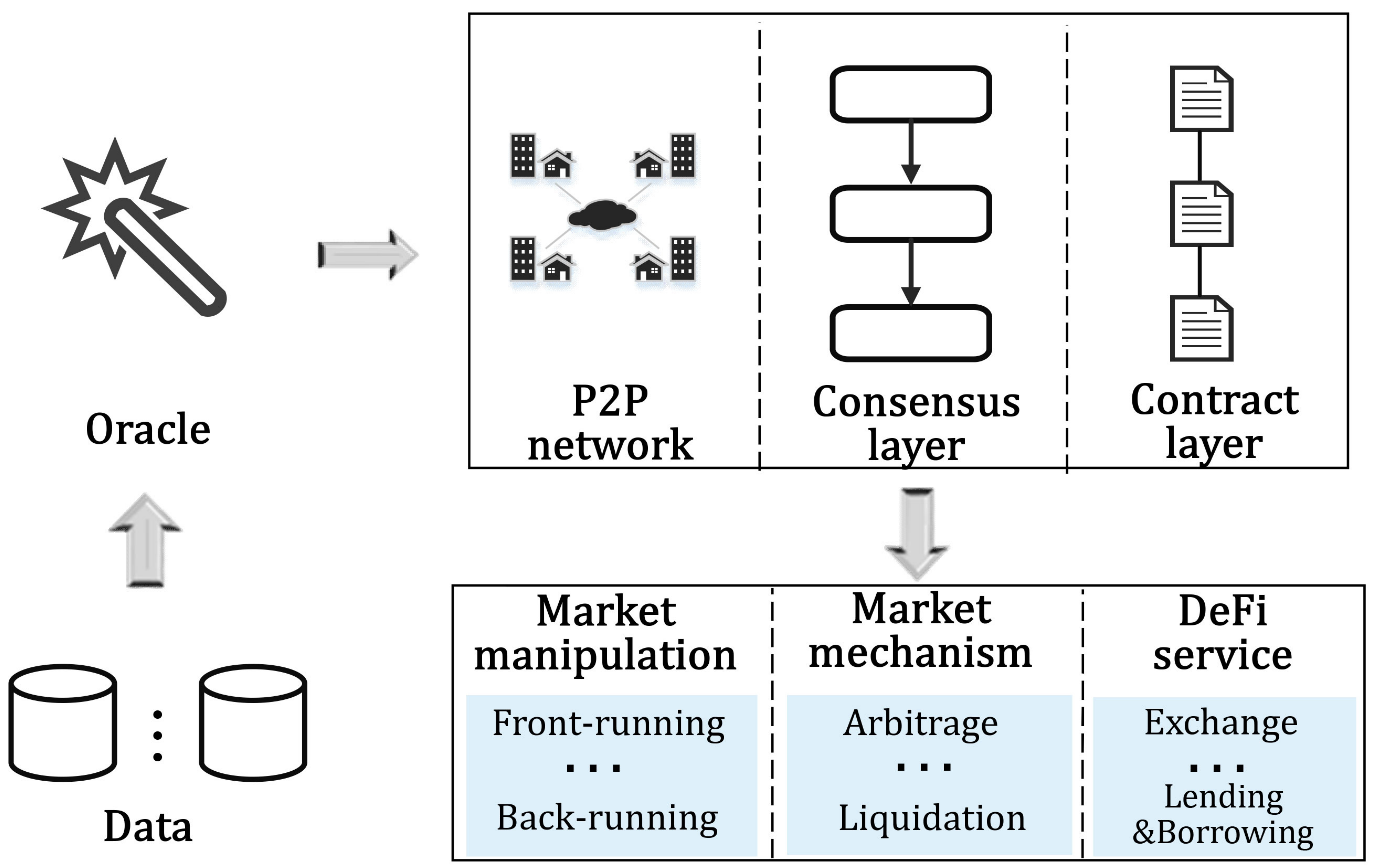}
	\caption{Overview of Research Ideas and Analysis Paths on DeFi Security.}
	\vspace{-4ex}
	\label{fig:overview}
\end{figure}

Moreover, the blockchain serves as the foundation of the DeFi application and enables transactions on DeFi to be completed securely. Blockchain's consensus mechanism ensures the integrity of DeFi transactions. The consensus mechanism selects the ledger nodes for the blockchain. The nodes with bookkeeping rights incorporate the DeFi application's transactions into a new block. The proper execution of the financial logic of the DeFi application relies on smart contracts~\citep{jensen2021introduction}. The smart contract isolates from the outside world and cannot be modified once deployed on the blockchain. In detail, to get reliable real-world asset price information, DeFi introduces the oracle~\citep{RN47}, which is a system to provide real-world financial asset price information. 

\begin{table*}[width=18cm, pos=ht]
\setlength{\abovecaptionskip}{0cm} 
\setlength{\belowcaptionskip}{-0.2cm}
\centering
\caption{Comparison of Our Study and Other DeFi Security Related Literature.}
\footnotesize
\renewcommand{\arraystretch}{1.3} 
\begin{tabular}{|p{2cm}|p{9.5cm}|p{1.5cm}|p{2cm}| } 

\ChangeRT{0.7pt}
       \textbf{Reference} & \textbf{Contributions} & \textbf{Date} & \textbf{Categories}   \\
\ChangeRT{0.05pt}
   \multirow{2}{2cm}{\cite{jensen2021introduction}} & \multirow{2}{9.4cm}{It focuses on the analysis of financial services. It classifies the risks of users, liquidity providers, arbitrageurs and application designers separately.} & \multirow{2}{2cm}{April \\2021}  & \multirow{2}{2.3cm}{Financial Risk}\\
  & & &  \\ 
\ChangeRT{0.5pt}
  \multirow{2}{2cm}{\cite{RN47}} & \multirow{2}{9.5cm}{It focuses on the economic aspects and classifies the financial risks encountered by DeFi. And it analyzes the DeFi protocol and ecosystem.}  & \multirow{2}{2cm}{September\\2021}  &\multirow{2}{2.3cm}{Financial Risk} \\ 
  & & &  \\
\ChangeRT{0.05pt}
  \multirow{2}{2cm}{\cite{RN61} }& \multirow{2}{9.5cm}{It first introduces the breadth of the lending market (a DeFi service). It quantifies the instability of lending protocols.}  & \multirow{2}{2cm}{November\\2021 } &\multirow{2}{2.3cm}{Financial Risk} \\
  & & & \\

\ChangeRT{0.05pt}
   \multirow{2}{2.5cm}{\cite{RN51}} & \multirow{2}{9.5cm}{It introduces a new type of Flash Loan attack and demonstrates the weaknesses and price fluctuations of the DeFi protocol.} & \multirow{2}{2cm}{June\\2020}  &\multirow{2}{2.3cm}{Technical Risk} \\
  & & &  \\   

\ChangeRT{0.05pt}
 \multirow{2}{1.5cm}{\cite{qin2021cefi} } & \multirow{2}{9.5cm}{It compares the differences between traditional CeFi and DeFi, including legal, economic, and security.} & \multirow{2}{2cm}{June\\2021 } &\multirow{2}{2.3cm}{Financial Risk}\\
 & & &  \\ 

\ChangeRT{0.05pt}
  \multirow{2}{1.5cm}{\cite{amler2021defi} } & \multirow{2}{9.5cm}{It classifies DeFi services through the economics dimension, highlighting the advantages of DeFi compared to traditional finance.} & \multirow{2}{2cm}{September\\2021}  & \multirow{2}{2.3cm}{Financial Risk}\\
  & & &  \\
  
\ChangeRT{0.05pt}
  \multirow{2}{2cm}{\cite{bartoletti2021towards} } & \multirow{2}{9.5cm}{It formalizes DeFi theory in order to analyze various DeFi incentive mechanisms and design principles.} & \multirow{2}{2cm}{September\\2021}  &  \multirow{2}{2.3cm}{Technical Optimization}\\
  & & &  \\ 

\ChangeRT{0.05pt}
  \multirow{2}{1.5cm}{\cite{liu2020mover}} & \multirow{2}{9.5cm}{Markov Chain and volatility prediction risk management are proposed. Loss distribution reduces mortgage rates, and VaR calculates external risks.} & \multirow{2}{2cm}{October\\2020}  & \multirow{2}{2cm}{Technical Optimization} \\
  & & & \\
\ChangeRT{0.05pt}
  \multirow{2}{1.5cm}{\cite{RN90} } & \multirow{2}{9.5cm}{It proposes a DeFi attack detection system that collects and analyzes transactions using symbol execution and transaction monitoring.} & \multirow{2}{2cm}{March\\2021} & \multirow{2}{2cm}{Technical Optimization} \\
  & & &  \\
\ChangeRT{0.05pt}
  \multirow{2}{1.5cm}{\cite{bekemeier2021deceptive}} &\multirow{2}{9cm}{It presents the first systematic risk and is the first empirical guide to stylized facts both at the technical level and economic level.} & \multirow{2}{2cm}{December\\2021}  & \multirow{2}{2cm}{Technical and Financial Risk} \\
  & & &\\
\ChangeRT{0.05pt}
  \multirow{2}{2cm}{Our study} & \multirow{2}{9.5cm}{It is the first to provide a systematic summary of DeFi security incidents and systematically analyze the vulnerabilities. We also provide future directions.}  &  \multirow{2}{2cm}{——} & \multirow{2}{2cm}{Systematic Review}\\
  & & & \\ 

\ChangeRT{0.7pt}
  \end{tabular}
  \vspace{-2ex}
  \label{tab:related_work}
\end{table*}

With the rapid development of DeFi, it can be divided into stablecoin, Decentralized Exchange (DEX), cryptocurrency market, and insurance. Additionally, it had locked in \$200 billion until April 2022~\citep{defillama}. However, the value locked up in the entire DeFi dropped by around \$85 billion in July 2022, causing us to ponder the security of DeFi.

While some studies about the risk of DeFi are in Table \ref{tab:related_work}, they paid more attention to financial issues. \cite{RN47} classified attacks according to risk categories from an economic perspective. \cite{RN61} systematically and quantitatively compared various lending systems and measured the risks that participants may encounter. \cite{RN51} described the design flaws in lending protocols and DeFi losses due to price volatility. \cite{qin2021cefi} systematically compared Centralized Finance (CeFi) and DeFi, including legal, economic, and market. \cite{bartoletti2021towards} formalized the DeFi theory, which was used to understand systematically and analyze the incentives in DeFi to balance interest rates and prices. Other studies proposed by \cite{jensen2021introduction} and \cite{amler2021defi} were used to analyze the risk of assets in DeFi on Blockchain.

In addition to the research of financial risks in DeFi, optimization schemes were also widely studied, as shown in Table \ref{tab:related_work}. \cite{liu2020mover} used a mathematical-statistical approach to the market for four types of assets and clearing to construct MovER, a framework for controlling the risk of the system. \cite{RN90} proposed Blockeye, which constructed state dependencies from smart contracts and used the collected transactions to analyze whether it is subject to a DeFi attack. Even though there are some optimized solutions to vulnerabilities, attacks keep appearing, such as the Ronin Bridge incident~\citep{RN23}. 

Similar work to ours was proposed by \cite{bekemeier2021deceptive}, it discussed systemic risk, both at the technical level of the blockchain and the economic level, and provided experience analysis. The difference is that our work is more comprehensive. Our work in this paper systematically summarizes vulnerabilities at all technical levels, following the analytical path shown in Figure~\ref{fig:overview}. In addition, we analyze the attack events caused by the vulnerabilities. Most importantly, we also summarize the most state-of-the-art optimizations at each layer. Finally, we conclude with some challenges and possible future directions.

The main contributions of this paper are as follows:
\begin{itemize}

    \item [(i)] To the best of our knowledge, we conducted the \textit{first} systematic examination of the security issues of the DeFi ecosystem built on blockchain.

    \item [(ii)] We systematically summarize the vulnerabilities of the Ethereum-based DeFi system, investigate real-world attack events related to DeFi and classify them according to their vulnerability principles.

    \item [(iii)] We survey the security optimizations in DeFi from the system level and conclude the challenges to suggest future research directions in this area.
\end{itemize}

The rest of the paper is structured as follows. Section \ref{sec:bg} presents the background of the paper. In Section \ref{sec:ana_v}, we examine some vulnerabilities in DeFi, and in Section \ref{sec:attack_events}, we analyze real-world attacks. Section \ref{Security_Optimization} provides several security optimizations, while Section \ref{Challenges_Future} highlights DeFi's challenges and future directions. Finally, Section \ref{conclusion} concludes the paper.

\section{Background}
\label{sec:bg}
\setParDis
\subsection{Blockchain}
\setParDef

\begin{table*}[width=18cm, pos=ht]
\setlength{\abovecaptionskip}{0cm} 
    \centering
    \caption{Summarization of Different Blockchains in Top 525 Popular DeFi Applications.}
    \renewcommand{\arraystretch}{1.3} 
\footnotesize
    \begin{tabular}{|p{4cm}| p{3cm}| c| c| c| c| c| c|}
        \ChangeRT{0.7pt}
        \multirow{2}{*}{\textbf{Names}} & \multirow{2}{4cm}{\textbf{Supported Consensus Algorithms}} & \multirow{2}{1.5cm}{\textbf{EVM-Compatible}} & \multicolumn{4}{c|}{\textbf{Features}} & \multirow{2}{*}{\textbf{Ratio}} \\
        \cline{4-7}  
        &  & & Sec & TP & Sca & TC & \\
        \ChangeRT{0.05pt}
         \multirow{1}{4cm}{Ethereum~\citep{buterin2014next}} &\multirow{1}{3cm}{PoW \& PoS \& PoA} & \checkmark & \CIRCLE & \CIRCLE  & \RIGHTcircle & \CIRCLE & 38.3\% \\
         \ChangeRT{0.05pt}
         \multirow{1}{4cm}{BNB~\citep{bnbchain}} &\multirow{1}{3cm}{PoSA} & \checkmark & \LEFTcircle & \LEFTcircle & \LEFTcircle & \RIGHTcircle & 34.7\% \\
         \ChangeRT{0.05pt}
         \multirow{1}{4cm}{Polygon~\citep{kanani2021matic}} &\multirow{1}{3cm}{PoS} & \checkmark & \Circle & \LEFTcircle & \CIRCLE & \Circle & 12.4\% \\
         \ChangeRT{0.05pt}
         \multirow{1}{4cm}{TRON~\citep{liu2018TRON}} &\multirow{1}{3cm}{DPoS \& TPoS} &\checkmark  & \LEFTcircle & \CIRCLE & \LEFTcircle & \RIGHTcircle & 5.5\% \\
         \ChangeRT{0.05pt}
         \multirow{1}{4cm}{Tezos~\citep{Tezos14}} &\multirow{1}{4cm}{Emmy$^*$ \& Tenderbake} & $\times$ & \CIRCLE & \LEFTcircle & \Circle & \RIGHTcircle & 2.1\% \\

        \ChangeRT{0.7pt}
    \end{tabular}
    \vspace{-2ex}
    \label{tab:blockchains}
\end{table*}

Blockchain~\citep{li2020survey} is an advanced Peer-to-Peer (P2P) database system constructed using cryptography and consensus algorithms. According to its evolution, four distinct eras can be distinguished. Bitcoin, which represents the Blockchain 1.0 era, focuses primarily on decentralized and cryptocurrency properties. With the introduction of smart contract technology, blockchain technology has entered the 2.0 era, which is dominated by Ethereum. Integration of smart contracts accelerates the advancement of DeFi technology~\citep{zhang2022secure}. To improve the scalability of blockchain in the 3.0 era, blockchain platforms such as EOS permit real-time interaction between multiple chains. The 4.0 era is defined by the optimal integration of blockchain technology and conventional industry.  

In Table \ref{tab:blockchains}, we survey the 525 most popular DeFi applications by popularity~\citep{dappcom2022} on 9 September 2022. With the advent of various blockchains, there are more options for DeFi applications deployment, such as Binance Smart Chain (BSC or BNB) \citep{bnbchain}, Polygon \citep{kanani2021matic}, TRON \citep{liu2018TRON}, Tezos \citep{Tezos14}, and others. Moreover, we discovered that the Ethereum comprised the largest proportion, followed by the BNB chain. In addition, in the "Features" column, 'Sec', 'TP', 'Sca', and 'TC' represents Security, Transparency, Scalability, and Transaction Costs, respectively. They are ranged with index units expressed as \Circle, \LEFTcircle, \RIGHTcircle$ $ and \CIRCLE$ $ in order from weak to good. We found that these blockchain platforms account for over 90\% of the 525 DeFi applications, with the majority of them supporting Ethereum virtual machine (EVM). Furthermore, other platforms are optimized based on Ethereum, sacrificing security or transparency to optimize scalability and transaction costs. Therefore, we prioritize Ethereum during the analysis in the rest of the paper.

\subsubsection{Ethereum}
Ethereum is a public blockchain system initialized using the Proof-of-Work (PoW) consensus mechanism, in which miners fight for control of blocks using computing power in exchange for incentives~\citep{li2020survey}. However, it has subsequently shifted to the Proof-of-Stake (PoS) algorithm in September 2022, which is based on the quantity and age of stakes held~\citep{wahab2018survey}. It first uses the Turing-complete programming language Solidity, Vyper, and others to develop smart contracts~\citep{chen2020understanding,li2020stan}. On the Ethereum blockchain, anyone can deploy decentralized applications (DAPPs) that can interact with other network nodes. DeFi, an application that provides various financial services, is the current market leader and one of the most popular applications in the financial industry.


\subsubsection{Other Blockchains}
The BNB chain~\citep{bnbchain} is a blockchain compatible with Ethereum Virtual Machine (EVM) that employs the Proof of Stake Authority (PoSA) consensus algorithm. Polygon~\citep{kanani2021matic} is an extension chain for the Ethereum. It utilizes the PoS consensus algorithm to increase transaction speed and the ZK-Rollup zero-knowledge proof technology to ensure transaction security. TRON~\citep{liu2018TRON} developed a TRON Virtual Machine (TVM) that is fully compatible with Ethereum smart contracts. It uses the TPoS consensus algorithm that combines DPoS and PBFT to balance security and Scalability. Tezos~\citep{Tezos14} employs the Emmy$^*$ and Tenderbake protocol based on the PoS consensus algorithm. It employs formal verification techniques at the protocol and application layers to ensure security. Moreover, it scales utilizing a mechanism that can be upgraded.

\subsection{Consensus Mechanism of Blockchain}
The consensus mechanism is the basic technology of the blockchain, which ensures the blockchain's secure, stable, and efficient operation. At the same time, the consensus mechanism enables the "mistrustful" parties on Ethereum to complete the verification and confirmation of transactions. Researchers are continuously improving various consensus mechanisms such as PoW, PoS, and DPoS~\citep{lashkari2021comprehensive}. 

\cite{nakamoto2008bitcoin} proposed PoW to prevent double spending on cryptocurrencies. The core idea of PoW is to compete among nodes for the bookkeeping rights and rewards of each block through their computing power~\citep{mingxiao2017review}. All miner nodes in the network use the information in the previous block, such as previous block hash, timestamp, and nonce, to determine the next block. In PoW, miner nodes find the hash value by continuously trying random number nonce, which is difficult to calculate but simple to verify. 

\cite{mingxiao2017review} proposed PoS, whose core idea is that the greater the ownership of a node to a specific amount of cryptocurrency, the greater the equity of the node. In PoS, it filters nodes by calculating the number of currencies in the nodes as a percentage of the total currencies and the time of holding currencies. This approach starts by selecting nodes, and only then moves on to carry out arithmetic operations. As a result, a significant amount of computational resources are not wasted.

Initial implementation of the PoA~\citep{wood2016polkadot} consensus protocol occurred on the Ethereum test chain. It requires the same authentication nodes as the PoS consensus algorithm, but its authentication nodes must pass more stringent and complex standards. On this basis, consensus can be reached without communication; however, the identity of the verification node is visible, and the rate of node forgery is proportional to the shareholding ratio.

DPoS is an additional PoS~\citep{wood2016polkadot} consensus algorithm variant. Participants combine their assets into a single pool of assets. Participants vote for the verifier, who then executes the asset pool trade. Moreover, the previous block's representative cannot be applied to the next block. Unlike PoS consensus, DPoS elects the verifier by vote without regard to the proportion of account assets.

PoSA~\citep{posa} is the synthesis of PoS and PoA consensus. It specifies a fixed number of verifiers ranked by the number of assets they possess. These verifiers are publicly accessible on the blockchain and take turns validating blocks. For instance, the BSC chain has 21 verifiers, whereas Ethereum has over 70,000. It compromises security for the sake of speed.

\cite{castro1999practical} proposed the PBFT algorithm based on state machine copy replication, where transactions are modeled as state machines, and the state machines are replicated at different nodes of the blockchain. Each copy of the state machine preserves the state of the transaction. The state is changed only when the blockchain reaches consensus, ensuring the strong consistency of the blockchain in real-time.

TPoS~\citep{TPoSConsensus} incorporates PBFT to improve the decentralization of DPoS. The final number of witness seats is used as the threshold, and each block slot has a default number of witness seats. Once all participants meet the criterion, a consensus is reached and rewards are given.

Emmy$^*$~\citep{Tezos14} utilizes 5256 block slots and determines if a block can be generated based on its block priority. 
Similarly to TPoS, the number of block endorsements greater than the total number of block slots is also used as a consensus reference factor.

Tenderbake~\citep{Tenderbake2021} is an Emmy$^*$-optimized consensus mechanism that utilizes two votes to determine which blocks achieve consensus the fastest. It decreases the time required to generate a block and allows the vote to be validated only once, which increases the network's stability.

\subsection{Layers of DAPP on Blockchains}
DAPPs, like traditional software architectures, can be separated into six layers as follows~\citep{duan2022multiple}: 
(1) The data layer handles off-chain data before passing it on to the network layer. (2) The network layer is peer-to-peer, assuring network node autonomy. (3) The consensus layer guarantees that miners package network layer requests. (4) The incentive and consensus layers are interrelated, and the incentive layer ensures that miners do not behave maliciously. (5) The smart contract layer connects the consensus with application layers and exchanges data between them, and (6) the application layer binds the information from the smart contract layer and shows it to the user after processing.

\subsection{Transaction Process on Blockchain}
When a user interacts with the applications and begins a transaction request using the interfaces provided by the smart contract, the transaction request broadcasts to all nodes on the P2P network chain~\citep{li2021hybrid}. When the miner gets the request, it selects and packages the transaction into blocks. The miner adds blocks to the chain using the consensus algorithm and synchronizes them with all nodes on the network. Simultaneously, the smart contract changes the state variables depending on transaction data and visualizes them in the application.

\subsubsection{Geth}
Go-Ethereum (Geth) is an official Ethereum client implemented in the go programming language~\citep{go_ethereum2022}. It includes instructions for several tasks, such as creating an Ethereum private chain and interacting with the network environment.

\subsubsection{Gas}
To avoid the overuse of network resources, all transactions on Ethereum are paid a cost called \texttt{gas}, and the transaction fee equals the amounts of \texttt{gas} multiplied by \texttt{gasPrice} \citep{chen2017under,chen2020gaschecker}. The user who proposes transactions sets the \texttt{gasPrice}, and miners with high computing resources would conduct the transaction earlier if the \texttt{gasPrice} is high~\citep{chen2018towards}. There is also a concept called \texttt{gaslimit}, which is used to limit the maximum amount of \texttt{gas} that can be used for a transaction~\citep{chen2017adaptive}. It means that the maximum charge for a transaction is \texttt{gaslimit} multiplied by \texttt{gasPrice}.

\subsubsection{Maximal Extractable Value (MEV)}
Regarding blockchains with PoW, the MEV is the value that can be extracted by the miner. However, the maximum extractable value may make more sense for blockchains utilizing PoS or other consensus algorithms.

Miner extractable value refers to the profit miners make by performing a series of operations on the blocks they mine \citep{RN7}. For example, miners reorder transactions to optimize the initial ordering of transactions and earn additional Ordering Optimization (OO) fees~\citep{RN3}. And the phenomenon that miners sell priority in blocks to make users keep raising the cost of gas is called Priority Gas Auctions (PGA).

Maximal Extractable Value is the maximum value that the validator $V$ can extract by reordering, inserting, or not executing the transactions $T_{i,...,j}=\{t_i,...,t_j\}$ in the block. In addition, we assume that the balance in $V$ before the transaction is $b(s)$ and $b(s')$ is after the transaction. So the value obtained by sequential execution $EV(V,T_{i,...,j})$ equals $b(s')-b(s)$, and $R(T_{i,...,j})$ means the order of transactions is in full array. Thus the maximal extractable value $MEV$ can be defined as $MEV=max(EV(V,R(T_{i,...,j})))$.

\subsection{DeFi}
\setParDis
\subsubsection{Development of DeFi}
\setParDef
The introduction of blockchain technology~\citep{nakamoto2008bitcoin} has changed the traditional financial ecosystem. With the advent of Ethereum, smart contracts became the basis for the development and implementation of DeFi. Since the landing of MakerDAO in 2014, which is the first Ethereum-based DeFi project, several DeFi protocols have emerged to implement functions of traditional CeFi, such as lending platforms, exchanges, derivatives, and margin trading systems~\citep{wang2022speculative}. As liquidity mining mentioned in 2020, DeFi applications were pushed into high gear with the emergence of DEXs such as Compound, which were entirely managed by smart contracts. Asset Legos brings unlimited creativity to DeFi products. It means that a new financial product can be realized by combining the underlying DeFi protocols~\citep{popescu2020decentralized}. In 2022, regulated Decentralized Finance (rDeFi) becomes the new trend in DeFi development~\citep{rDeFi2022}.

\subsubsection{DeFi Service}
As depicted in Figure~\ref{fig:overview}, DeFi applications can be made up of DeFi services, also known as protocols, such as exchange, lending, and asset operation. Blockchain will wait for assets or data to be processed through protocols before uploading them to the application layer, which is the market ~\citep{schar2021decentralized}. The DEX serves as a forum for asset suppliers and buyers to engage. It can be separated into two types: centralized order system and Automated Market Maker (AMM)~\citep{RN2}. The former is comparable to a regular exchange in that customers produce trade orders following transactions start. The latter is accomplished quickly by initiating a transaction using a previously constructed asset price algorithm.
\subsubsection{Market Mechanism}
In addition to technological issues, DeFi has an economic mode of operation, which is the market mechanism. Users can control and alter numerous assets using the DeFi service normally. However, attackers can benefit by manipulating the asset through market-based strategies at the economic level. 

\section{Analysis of Vulnerabilities}
\label{sec:ana_v}
From the proposal of DeFi to 2022, various vulnerabilities have emerged to promote the ecological development of DAPPs. Therefore, studying the vulnerabilities associated with DeFi facilitates comprehension of attack defense techniques. In order to provide a concise summary of the dangers posed by DeFi, we will concentrate on the data, consensus, contract, and application layers.

\subsection{Data Security Vulnerabilities}
\label{sec:data_security_v}
\begin{figure}[ht]
\setlength{\abovecaptionskip}{0cm} 
\setlength{\belowcaptionskip}{-0.4cm}
    \centering
    \includegraphics[width=8cm]{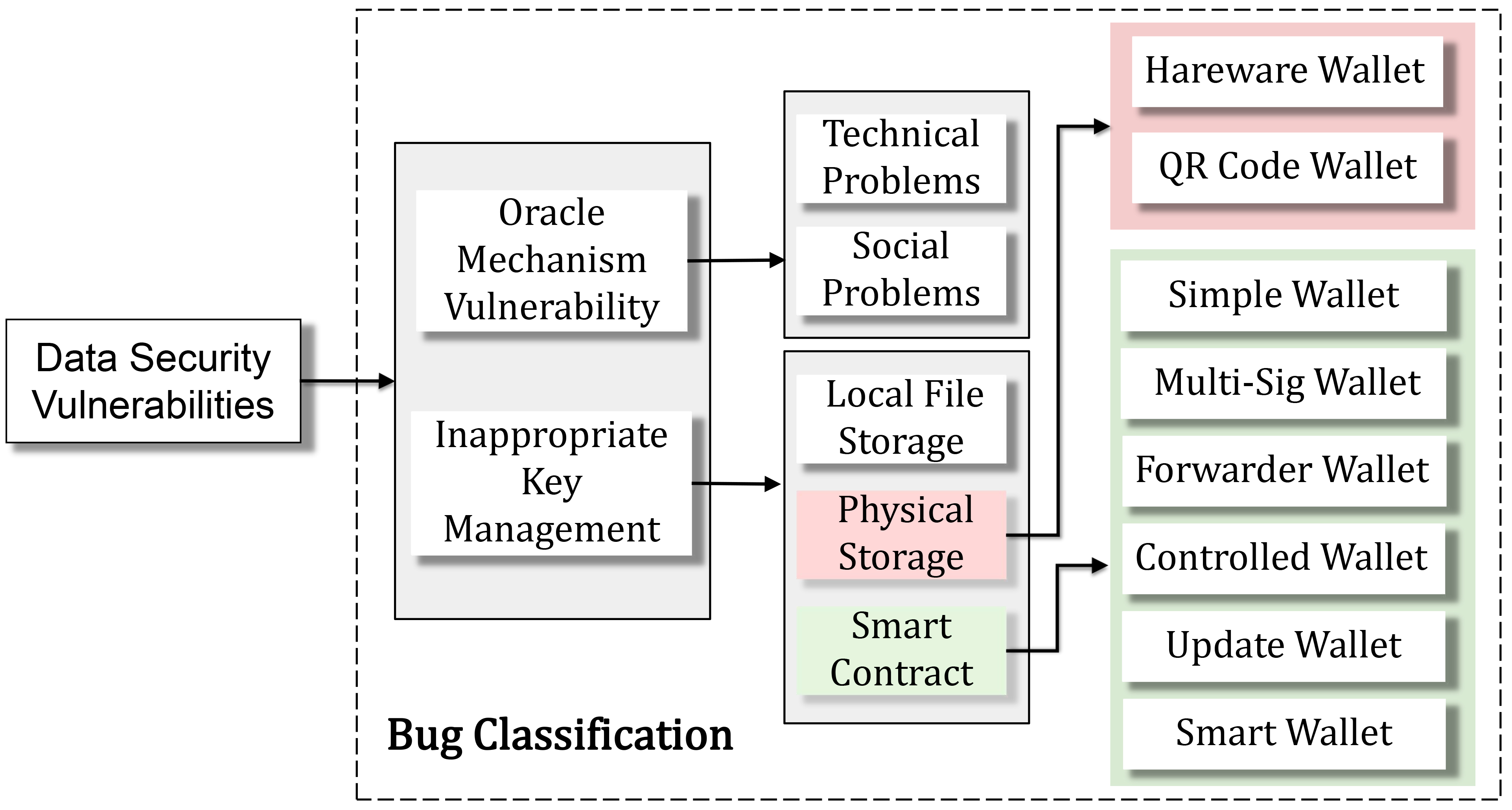}
    \caption{Classification of Data Security Vulnerabilities.}
    \vspace{-3ex}
    \label{fig:Class_data_security}
\end{figure}

For the data layer, if attackers change the data under the chain during the uploading process to the chain, it will result in irreversible mistakes due to the immutability of the blockchain. Figure~\ref{fig:Class_data_security} shows that it could encounter oracle mechanism vulnerability and inappropriate key management.

\subsubsection{Oracle Mechanism Vulnerability}
The oracle is an automated service mechanism that allows the system to obtain the off-chain asset price data as input~\citep{RN47}. And smart contracts rely on the exchange rates of prices provided by oracle for proper operation. However, as Figure~\ref{fig:overview} shows, the risk to oracle grows drastically when a single point of failure occurs. For example, over 3 million sETH were arbitrated due to the oracle errors in \textsc{Synthetix}, a protocol that converts entity into synthetic~\citep{RN53}. Oracle risks can be divided into technical and social problems.

Technical oracle problems may be defined as a process of passing data with three key elements: (1) How to collect all the data accurately? (2) How to process the data with as few errors as possible? And (3) How to upload the processed data to the smart contract?  

Furthermore, the current oracle form may be centralized and distributed. Centralized oracle uses trusted third parties to collect, process, and transfer data to smart contracts. Distributed oracle consists of numerous nodes that take data from multi-sources and process it using an algorithm, such as a consensus~\citep{kumar2020decentralising} or weighted voting method~\citep{angeris2020improved}. Finally, the oracle system assesses the chain information.

\begin{table*}[width=18cm, pos=ht]  
\setlength{\abovecaptionskip}{0cm} 
\setlength{\belowcaptionskip}{-0.2cm}
\centering  
\caption{Comparison of Different Key Storage Methods on Ethereum.}
\footnotesize  
\renewcommand{\arraystretch}{1.3} 
\begin{tabular}{|l| p{9.8cm}| c| c| c| c| c|} 
\ChangeRT{0.7pt}
\multirow{2}{*}{\textbf{Wallets}}& \multirow{2}{*}{\textbf{Descriptions}}&\multicolumn{5}{c|}{\textbf{Features}}   \\
\cline{3-7}  
  &  & Flex & Sec & Sca & TP & TxC \\
\ChangeRT{0.5pt}
  Local Storage & Keys are stored centrally in the file system by default  & -  & - & -  & - & - \\
\ChangeRT{0.05pt}
  Hardware Wallet & Hardware devices can isolate external networks and transport operations  & $\times$  & \checkmark & $\times$  & $\times$ & - \\
\ChangeRT{0.05pt}
  QR Code Wallet & QR code generated from the address and scanned to obtain the address  & $\times$  & \checkmark & $\times$  & $\times$ & - \\
\ChangeRT{0.05pt}
  Simple Wallet & It can simply handle cryptocurrencies and tokens for raw transactions  & \checkmark  & - & \checkmark  & - & \checkmark\\
\ChangeRT{0.05pt}
  Multi-Sig Wallet & The transaction process requires multiple owners to sign to ensure users' security  & $\times$  & \checkmark & \checkmark  & \checkmark & \checkmark\\
\ChangeRT{0.05pt}
  Forwarder Wallets & Forwarding assets to a master wallet and users only need to preserve the subkey  & \checkmark  & \checkmark & \checkmark  & $\times$ & \checkmark\\
\ChangeRT{0.05pt}
  Controlled Wallets & The third party keeps the key and anyone who uses the key needs authorization & $\times$  & \checkmark & $\times$  & $\times$ & \checkmark\\
\ChangeRT{0.05pt}
  Update Wallets & Users can customize the update by selecting some parts to be updated  & \checkmark  & \checkmark & \checkmark  & $\times$ & \checkmark\\
\ChangeRT{0.05pt}
  Smart Wallets & Wallets with enhanced functionality that achieve expansion of normal functions  & \checkmark  & $\times$ & \checkmark  & $\times$ & \checkmark\\
\ChangeRT{0.7pt}
  \end{tabular}
  \vspace{-1.5ex}
  \label{tab:key_storage}
\end{table*}

There are not only technical problems but also social problems in oracle~\citep{caldarelli2021blockchain, egberts2017oracle}. Assuming such a game where there exists an Oracle $O_i$. The $O_i$ picks the off-chain data and processes it as $D_n=(d_i,...,d_j)$. The contract $S_i=(f_{i1},...,f_{im})$ uses $D_n$ for transactions $Tx_i$, where $f_{ii}$ is the $i_{th}$ function in the contract $S_i$. If an attacker $a_i$ pays $c$ to modify $d_i$ in $D_n$, and obtains benefits $b_{ii}$. When the cost $c$ by the attacker is less than the benefits $b_{ii}$, the attacker gets a profit that would be attractive to other attackers. While the $c$ cannot be measured directly from technical methods, it requires analysis of specific social situations, so the oracle problem is controversial in terms of social issues.

\subsubsection{Inappropriate Key Management}
In the DeFi ecosystem, wallets are used to manage private keys, and asset authentication is based on keys in most cases. However, similar to Bitcoin, the DeFi system suffers from the problem of improper key management. Existing key management methods, such as physical storage \citep{RN64,shbair2021hsm}, offline wallets~\citep{RN80,RN74}, and password-derived wallets~\citep{kaliski2000pkcs}, have some drawbacks. In Table \ref{tab:key_storage}, we summarize nine forms of wallets, where local storage is the initial form of local file storage, hardware wallets and QR code wallets both belong to physical storage wallets. The remains belong to smart contract wallets. Moreover, 'Flex' in Table \ref{tab:key_storage} is the Flexibility, above the Local Storage is a $\checkmark$, and vice versa is a $\times$. The same applies to 'Sec', 'Sca', 'TP', and 'TxC', representing Security, Scalability, Transparency, and Transaction Costs, respectively. 

In Ethereum, users can access the Ethereum chain by using Geth. When a user creates an account $a_i$, the client generates a file to be stored locally, which contains the unique key $key_i$ associated with the account $a_i$. Before the account initiates a transaction $Tx_i$ or mining, the client reads the $key_i$ in the file. However, anyone without restricted access can read the file and even falsify $(key_i,...,key_j)$ for profit.

There are three types of wallets, software, hardware, and paper, depending on the form in which they exist~\citep{suratkar2020cryptocurrency}. Hardware and paper-based storage, which are physical storage, are more secure because they store keys in a way that isolates them from multi-user interaction. Nevertheless, it also has the weaknesses of poor scalability~\citep{arapinis2019formal} and the inability to have a single point of failure caused by the architecture design~\citep{RN64}.


Smart contract wallets are divided into six types in ~\citep{di2020characteristics}. They restrict the direct access to assets and provide some Application Binary Interfaces (ABIs) for manipulating data.

\textbf{Simple Wallet.} It is the initial form of wallet, offering simply raw transaction capability and storing all keys in files. When a malicious parity obtains file system permissions, keys can be read or even manipulated.

\textbf{Multi-signature Wallet.} It requires the co-signature of many owners for increased protection. The combination of many signatures dilutes the individual's influence, providing decentralization. And the public multiple signature combination could enhance transparency.

\textbf{Forwarder Wallet.} It adds forwarding operations to the signing process, such as password-derived wallets, which allow users to customize the master key and then derive sub-keys from controlling the asset. The forwarding operation faces a balance between transparency and security. If the derivation algorithm is publicly available, attackers who got the master key in some ways will reproduce the derivation process to obtain all sub-keys.

\textbf{Controlled Wallet.} The custodial wallet is an example of a controlled wallet since it keeps ownership of the account and grants access to users. It offers some protection by centralized management, but the non-transparent action also tests managers' credibility.

\textbf{Update Wallet.} Update wallets permit users to modify updates depending on features, allowing for greater flexibility in wallet operation. However, compatibility across many versions might result in worse security.

\textbf{Smart Wallet.} Smart wallets include some sophisticated features, such as key recovery. As a result, the smart contract enables wallets to execute various services in addition to transferring money. However, it adds to the dangers involved with smart contracts in Section \ref{sec:smart_contract_v}.

\begin{table*}[width=18cm, pos=ht]
\setlength{\abovecaptionskip}{0cm} 
\setlength{\belowcaptionskip}{-0.2cm}
\centering
 \caption{Summarization of Consensus Vulnerabilities on Geth that have Endangered DeFi.}
\footnotesize
\renewcommand{\arraystretch}{1.3} 
\begin{tabular}{|p{3cm}| p{9cm}| p{2cm}| l|} 
\ChangeRT{0.7pt}
       \textbf{Brief Explanation} & \textbf{Descriptions} & \textbf{Date} & \textbf{Severity}   \\
\ChangeRT{0.5pt}
  Journaling Mechanism & Geth can't restore a deleted empty account due to out-of-gas & November 2016 & High   \\
\ChangeRT{0.05pt}
   EVM Stack Underflow & SWAP, DUP, and BALANCE underflow the EVM stack & February 2017 & High  \\
\ChangeRT{0.05pt}
  Stack Elements & In a static environment with fewer than three stack elements & October 2017 & Low\\
\ChangeRT{0.05pt}
  Encryption Algorithm & The elliptic curve algorithm was not fully validated & February 2018 & High  \\
\ChangeRT{0.05pt}
  Timestamp Overflow & Timestamp, state variables in blocks, overflow & March 2019 & High \\
\ChangeRT{0.05pt}
  Shallow Copy & Pre-compile contract, making Geth inconsistent with memory & July 2020 & High \\
\ChangeRT{0.05pt}
  Ether Shift & Transfering the balance of the deleted account to the new account & August 2020 & High \\
\ChangeRT{0.05pt}
  Certain Sequences & Certain transaction sequences can lead to the failure of consensus & December 2020 & Middle \\
\ChangeRT{0.05pt}
  Incorrect Requirements & Failure to properly authorize timestamp leads to double spending & February 2021 & High \\
\ChangeRT{0.05pt}
  Memory Corruption & RETURNDATA corruption due to data replication, resulting in forking & August 2021 & High \\
\ChangeRT{0.05pt}
  Denial of Service (DoS) & Combination of short-term restructuring and delayed consensus decision  & October 2021 & Critical \\
\ChangeRT{0.05pt}
  Bignum Overflow & Some large values in consensus specification overflow leads to a fork & April 2022 & High \\

\ChangeRT{0.7pt}
  \end{tabular}
  \label{tab:consensus_bugs}
\end{table*}

\subsection{Consensus Mechanism Vulnerabilities}
\label{sec:con_mec_v}
\begin{figure}[ht]
\setlength{\abovecaptionskip}{0cm} 
\setlength{\belowcaptionskip}{0.2cm}
    \centering
    \includegraphics[width=7.5cm]{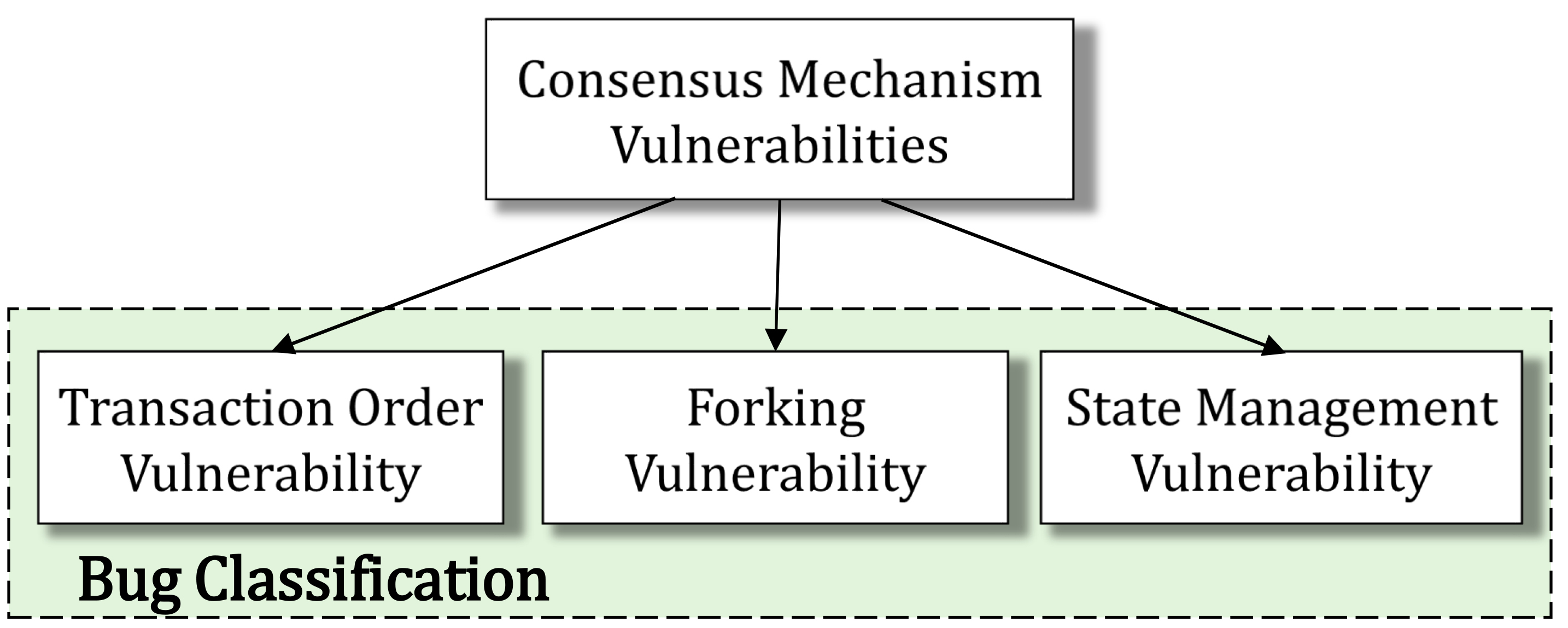}
    \caption{Classification of Consensus Mechanism Vulnerabilities}
    \vspace{-2ex}
    \label{fig:Class_Consensus_vul}
\end{figure}

Blockchains, such as Ethereum and others, are consensus-based. Up to now, many significant works have already been done in design, testing, auditing, and maintenance. So there are not many consensus flaws, but we gather the consensus bugs that occurred in Geth according to ~\citep{national2022consensus,yang2021finding,luu2015demystifying} in Table \ref{tab:consensus_bugs}, and we classify them from three aspects in Figure~\ref{fig:Class_Consensus_vul}. There are four severity categories, with low suggesting that the developer resolved before they occur. The 'Middle' was deployed to the test network before being discovered, while the ‘High’ was in the main chain. The 'Critical' implied that the vulnerability was widely available and had a significant impact on the integrity of the network.

Certain malicious behaviors utilize consensus rules to affect the sequences of transactions. There are a variety of attacks combined with MEV, such as flash loans~\citep{qin2021attacking,RN1}, sandwich attacks~\citep{RN2,RN7}, and forking attacks~\citep{RN3}. As Figure \ref{fig:Class_Consensus_vul} depicts, we classify this part into three segments: 1) Transaction Order Vulnerability, 2) Forking Vulnerability, 3) State Management Vulnerability.

\subsubsection{Transaction Order Vulnerability}
Transaction order vulnerability describes that an attacker alters the initial sequence of transactions by leveraging the miner's desire for profit. The sandwich attack~\citep{RN2} is a typical example. The attacker predicts that the victim will buy asset A, and pays a higher gas fee to acquire it before the victim at a lower price. And then, they sell A at a higher price for arbitrage since the victim's purchase boosts the price. 

\textbf{Front Running Attack.}
A front-running attack~\citep{RN2} employs a high mining cost to induce miners to block the original block and give the trade of attackers priority over the original trade, thereby modifying the status information of the actual trade in order to profit. In one instance, an attacker configures a bot to monitor profitable trades automatically and then attacks them automatically. However, as more individuals attack the same transaction, the potential for profit from each attack diminishes.

\textbf{Sandwich Attack.}
The sandwich attack~\citep{RN2} is a variant of the front run attack that employs both a front running attack and an end running attack, in which a second trade is made after the initial trade to maximize the benefit gained from the initial trade. In some DEXs, trades between decentralized market makers are automated. Trades have acceptable price slippage. When the price slip threshold is reached, the transaction is canceled. However, sandwich attacks permit the victim's trade to return to the slippage point. Moreover, we explain how it attacks in detail in Section \ref{subsec:sandwich_attack}.

\subsubsection{Forking Vulnerability}
Forking events in DeFi are generally associated with transaction fee-based forks and time-bandit attacks~\citep{RN3}. Mining revenue incentivizes miners to perform honestly, but the OO fee motivates them to reorder transactions in the block, enhancing the income. There are already many classical fork attacks in blockchain, including 51\% attacks \citep{51-attack} and selfish mining~\citep{kkedziora2019analysis} based on PoW, and Nothing at Stake attacks~\citep{Nothing-at-stake2022} based on PoS consensus. Most bugs contain forking vulnerabilities in Table \ref{tab:consensus_bugs}, for example, memory corruption, incorrect requirements, shallow copy, and certain sequences.

\textbf{51\% Attack.}
The 51\% attack~\citep{51-attack} occurs when attackers take control of the block generation. They possess the majority of hash computing power, which is the ability to solve difficult problems in a PoW consensus-based blockchain network. Typically, it can be separated into the direct profit and the indirect profit. The former is a double-spending attack in which the user's transaction is paid twice, doubling the profit. The later refers to DoS attacks that block transactions to compel other miners to join their pool, thereby increasing computing power and making it easier to generate a profit. Moreover, we have summarized some typical 51\% attack cases in Table \ref{tab:realworld_attacks1}.

\textbf{Selfish Mining.}
Selfish mining~\citep{kkedziora2019analysis} is that miners do not disclose the first mined block and then disclose two blocks after digging the second block, resulting in a fork of the current blockchain network. The advantage of this strategy is that one block has a high level of dominance over the others, which makes getting two blocks in a row more advantageous.
In order to perform this attack, the adversary must maintain two blockchains. The first is public for connecting to the mainnet chain, while the second is private for mining a second block. Once the first block is mined, it starts mining the second block. Even if other mines in the mainnet chain extract the block, the attacker can open their own mined block, possessing the priority to extract the second block based on the first height.


\textbf{Nothing at Stake.}
Nothing at stake~\citep{Nothing-at-stake2022} implies that, after the system used the PoS consensus algorithm has reached consensus with a block, it discovers that the block contains no information but still causes a fork. Even though the block reaches consensus with the chain, it takes the stake to another fork. The PoS consensus algorithm determines that the specified extractor generated the block if $(N*\$*t)$ is greater than the threshold. Where $n$ represents the random number generated by the extractor ID and public key, \$ represents the number of assets in the system, and $t$ represents the time of the extractor on the block that cannot be determined. Time and number of assets as common factors in reaching consensus, so attackers can use the little stake to reach consensus in many forks.

\subsubsection{State Management Vulnerability}
Transactions in Ethereum are based on updating states between blocks~\citep{RN30}. According to the consensus rules, the confirmation between the old and new blocks needs to be completed within certain minutes. Therefore, if attackers complete the extraction of the state variables within the block, then they can attack the transaction within the specified time. For example, timestamp overflow and incorrect requirements are in Table \ref{tab:consensus_bugs}. The former is because the timestamp exceeds the representation of \texttt{uint64}, resulting in a hash error in the block~\citep{yang2021finding}. The latter is that the timestamp in a block gets the permission mistake, which means the block is to be refused by the chain permanently, causing a chain fork and the execution of a double-spending attack~\citep{national2022consensus}. 

\textbf{Timestamp Overflow.}
An attacker uses an operation, such as uint64 in Ethereum, to make a timestamp attribute in a block exceed what it can represent. The false representation of the timestamp causes the block's hash value to change, preventing it from agreeing with the other nodes in the blockchain. In addition, the attack is typically caused by poorly designed modules that can obtain the blockchain's state. Moreover, it is available in Table \ref{tab:smart_contract_vulner} about smart contracts vulnerabilities.

\textbf{Incorrect Requirement.}
Incorrect Requirement refers to the attacker exploiting a configuration error in the server node timestamp information to render the block authentication invalid, resulting in node operation deformation. In some blockchains, network nodes are required to authorize the timestamp information for their nodes. However, there are some special conditions exist, such as the server setting the incorrect time zone or the network latency between network nodes.

\subsubsection{Others Consensus Attacks}
\textbf{Block Discard.} Block Discard is also called "Block Withholding". It indicates the phenomenon in mining pools where malicious miners send partial proofs of work to the pool manager and discard the full proofs of work~\citep{rosenfeld2011analysis}. This attack is commonly seen in competition between different mining pools.

\textbf{Pool Hopping Attack.} After analyzing the revenue of multiple mining pools, the attacker will choose to join the one with the highest revenue to join and allocate a portion of the arithmetic power to prevent it from mining blocks~\citep{rosenfeld2011analysis}. Similarly, when the mining pool manager finds that a certain blockchain network can get higher revenue, he will transfer the computing power to that network. However, the revenue of miners in the pool is still distributed according to the original revenue. The mining pool management can get more revenue. 

\textbf{Distributed Denial of Service Attack.}
Distributed Denial of Service (DDOS) attackers sever connections between multiple nodes and the network, thereby impacting the availability of the network or system. Regarding blockchain platforms, a DDoS attack against multiple nodes has a greater negative impact on DeFi than a DoS attack against a single node~\citep{singh2020utilization}. Additionally, DDoS attacks deplete the victim's computing and communication resources in a short period, slowing down the consensus agreement.

\textbf{Eclipse Attack.}
When all other nodes connected to the current node are attacked, the current node is under eclipse attack~\citep{RN105}. The attack prevents the P2P network's nodes from controlling information access. When all the nodes connected to the current node are the attacking node, the current node has suffered an eclipse attack. The attack prevents the P2P network's nodes from controlling information access. In reality, not all network nodes are interconnected, meaning that the attacker who takes control of a subset of the network can launch an eclipse attack. Furthermore, it can benefit from combined with a double spending attack~\citep{RN102}.

\textbf{Sybil Attack.}
The Sybil attack and the eclipse attack are similar, but the target of the attack is different. The eclipse attack intends to control the information in a specific node. In contrast, the Sybil attack prefers to attack all the nodes, reducing the backup function of the network. The Sybil attacker masquerades as a node participating in the election to deceive other honest nodes. It misleads other nodes by sending messages using fake identities to determine the connection status of blockchain \citep{bhutta2021survey}. When a certain number of fake nodes affects the blockchain's consensus, combining Sybil and 51\% attacks increases the ratio of a successful double spending attack. Sybil attacks are available in two modes, 1) the direct attack, where honest nodes are affected directly by fake nodes, and 2) the indirect attack, where honest nodes are impacted by nodes that communicate with Sybil nodes. Sybil attacks \citep{sybilevent} also occur in blockchains, where attackers create multiple fake network identities to inflate the value of the Sabre protocol and the Solana blockchain.

\subsection{Smart Contract Vulnerabilities}
\label{sec:smart_contract_v}
There are 20 types of smart contract vulnerabilities in blockchain defined in~\citep{RN34}, of which Table \ref{tab:smart_contract_vulner} shows the weaknesses that attackers might use to make a profit. We searched Common Vulnerabilities \& Exposures (CVE) and summarized over 500 vulnerabilities~\citep{CVE2022}. In Figure~\ref{fig:Class_Smartcontract_Application_vul}, we describe the classification of smart contract vulnerabilities in this paper. And Table~\ref{tab:smart_contract_vulner} shows that bugs written by Solidity were categorized into several types as detailed below:

\begin{figure}[ht]
\setlength{\abovecaptionskip}{0cm} 
\setlength{\belowcaptionskip}{-0.2cm}
    \centering
    \includegraphics[width=8cm]{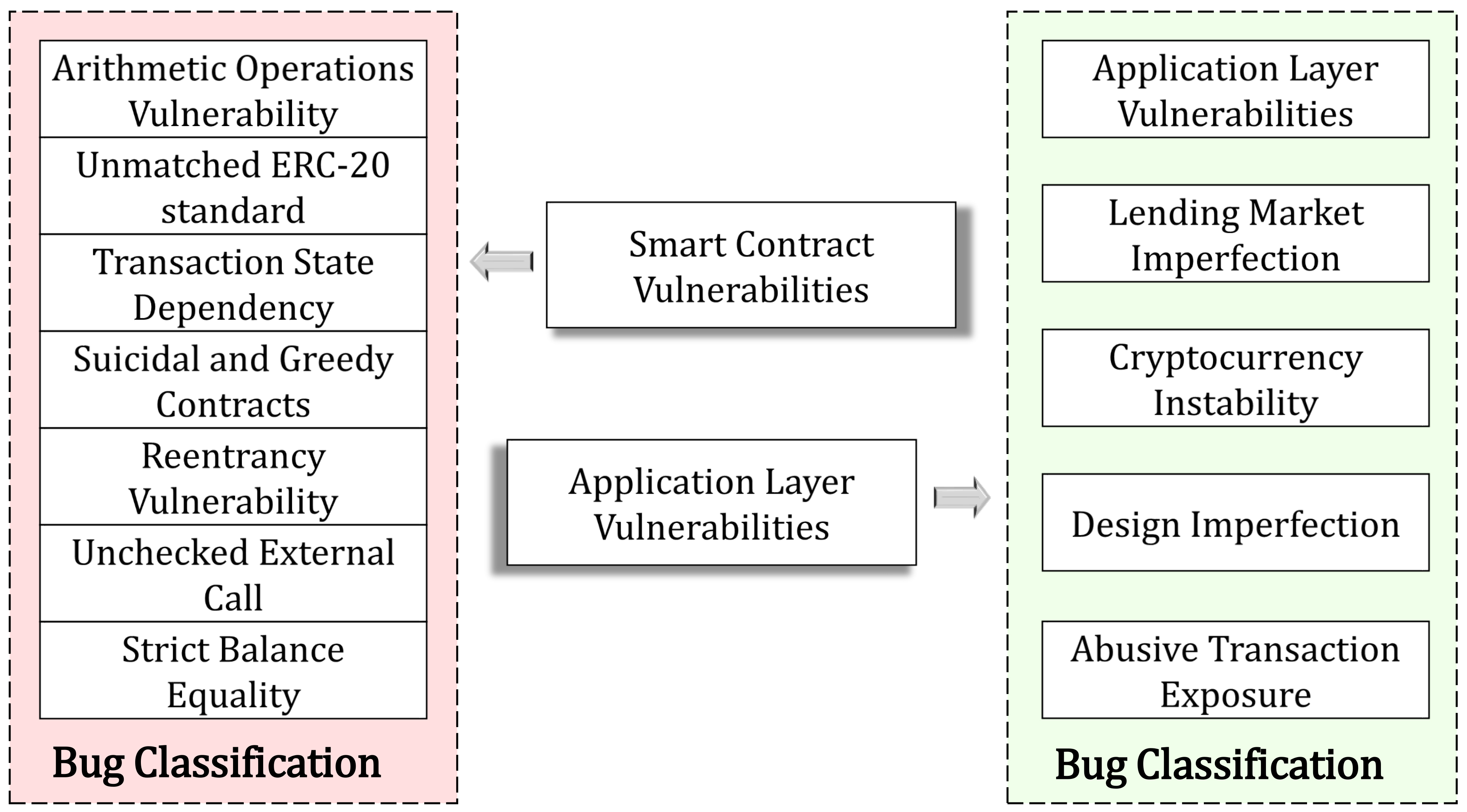}
    \caption{Classification of Smart Contract and Application Layer Vulnerabilities}
    \vspace{-4ex}
    \label{fig:Class_Smartcontract_Application_vul}
\end{figure}

\begin{table*}[width=18cm, pos=ht]
\setlength{\abovecaptionskip}{0cm} 
\setlength{\belowcaptionskip}{-0.2cm}
    \centering
    \caption{Summarization of Smart Contract Vulnerabilities in DeFi}
\footnotesize
    \renewcommand{\arraystretch}{1.3} 
    \begin{tabular}{|p{4cm}| p{4cm}|| p{3cm}| p{4.5cm}|}  
        \ChangeRT{0.7pt}
        \textbf{Categories} & \textbf{Causes} & \textbf{Categories} & \textbf{Causes}\\
        \ChangeRT{0.2pt}
        Unchecked External Calls & Without checking return values &Reentrancy&Repeated calls before completed  \\
\ChangeRT{0.05pt}
        Unexpected Permission Check&Failure to check permissions& Nested Call& Unrestricted call depth \\
\ChangeRT{0.05pt}        
        DoS Under External Influence&External exceptions inside loops& Missing Return&Denote return but no value \\
\ChangeRT{0.05pt}
        Unmatched ERC-20 Standard&Not follow the standard&Greedy Contracts & Receive but do not withdraw Ethers\\
\ChangeRT{0.05pt}
        Strict Balance Equality & Balance check failed & Block Info Dependency& Status in blocks leakage  \\
\ChangeRT{0.05pt}
        Misleading Data Location & Incorrect \texttt{storage} type & Missing Interrupter & No backdoor to handle crises\\
\ChangeRT{0.05pt}
        Transaction State Dependency & Error using \texttt{tx.origin} & Arithmetic Bugs&Unmatched type to values\\

        \ChangeRT{0.7pt}
    \end{tabular}
    \vspace{-2ex}
    \label{tab:smart_contract_vulner}
\end{table*}

\subsubsection{Arithmetic Operations Vulnerability}
There are a wide variety of bugs affecting the Solidity programming language. Common arithmetic manipulation bug includes integer overflow, float lack of precision, and division by zero.

An upward overflow can occur if a \texttt{memory} integer exceeds the maximum range, e.g., \texttt{uint256} is a default type of integer that can express the number from $0$ to $2^{256}-1$. In Listing \ref{lst:Integer}, the function allows the \texttt{owner} to add tokens to the user, but a sufficient \texttt{amount} on line 3 can make the balance in \texttt{balance[target]} vanish.



\definecolor{verylightgray}{rgb}{.97,.97,.97}

\lstdefinelanguage{Solidity}{
	keywords=[1]{anonymous, assembly, assert, balance, break, call, callcode, case, catch, class, constant, continue, constructor, contract, debugger, default, delegatecall, delete, do, else, emit, event, experimental, export, external, false, finally, for, function, gas, if, implements, import, in, indexed, instanceof, interface, internal, is, length, library, log0, log1, log2, log3, log4, memory, modifier, new, payable, pragma, private, protected, public, pure, push, require, return, returns, revert, selfdestruct, send, solidity, storage, struct, suicide, super, switch, then, this, throw, transfer, true, try, typeof, using, value, view, while, with, addmod, ecrecover, keccak256, mulmod, ripemd160, sha256, sha3}, 
	keywordstyle=[1]\color{blue}\bfseries,
	keywords=[2]{address, bool, byte, bytes, bytes1, bytes2, bytes3, bytes4, bytes5, bytes6, bytes7, bytes8, bytes9, bytes10, bytes11, bytes12, bytes13, bytes14, bytes15, bytes16, bytes17, bytes18, bytes19, bytes20, bytes21, bytes22, bytes23, bytes24, bytes25, bytes26, bytes27, bytes28, bytes29, bytes30, bytes31, bytes32, enum, int, int8, int16, int24, int32, int40, int48, int56, int64, int72, int80, int88, int96, int104, int112, int120, int128, int136, int144, int152, int160, int168, int176, int184, int192, int200, int208, int216, int224, int232, int240, int248, int256, mapping, string, uint, uint8, uint16, uint24, uint32, uint40, uint48, uint56, uint64, uint72, uint80, uint88, uint96, uint104, uint112, uint120, uint128, uint136, uint144, uint152, uint160, uint168, uint176, uint184, uint192, uint200, uint208, uint216, uint224, uint232, uint240, uint248, uint256, var, void, ether, finney, szabo, wei, days, hours, minutes, seconds, weeks, years},	
	keywordstyle=[2]\color{teal}\bfseries,
	keywords=[3]{block, blockhash, coinbase, difficulty, gaslimit, number, timestamp, msg, data, gas, sender, sig, value, now, tx, gasprice, origin},	
	keywordstyle=[3]\color{violet}\bfseries,
	identifierstyle=\color{black},
	sensitive=false,
	comment=[l]{//},
	morecomment=[s]{/*}{*/},
	commentstyle=\color{gray}\ttfamily,
	stringstyle=\color{red}\ttfamily,
	morestring=[b]',
	morestring=[b]"
}

\lstset{
	language=Solidity,
	extendedchars=true,
	basicstyle=\normalsize\ttfamily,
	showstringspaces=false,
	showspaces=false,
	numbers=left,
	numberstyle=\normalsize,
	numbersep=9pt,
	tabsize=2,
	breaklines=true,
	showtabs=false,
	captionpos=b
}

\begin{lstlisting}[language=Solidity,
                    numbers=left,
                    captionpos=b,
                    % belowskip = -0.7em,
                    caption=Integer Overflow Instance,
                    label=lst:Integer,
                    % title=Listing 1:\textbf{ Integer overflow instance}
                    ] 
function mintToken(address target, uint256 amount) onlyOwner{
  require(target != 0x0);
  balance[target] += amount;
  totalSupply += amount;
  Transfer(0, this, amount);
  Transfer(this, target, amount); 
 }
\end{lstlisting}

Since Solidity lacks the float type of data structure, the phenomenon in which the float result of an operation might lose coins. When one integer is divided by a larger integer, the result is always 0. For example, 1 ETH divided by 10 Eth equals 0. Even some contracts do not restrict the operation of division by zero, which results in code logic errors as the result of the calculation becomes big infinitely.

\subsubsection{Unmatched ERC-20 Standard}
Ethereum provides various APIs for developers to implement certain functions, such as transferring money, but some developers may not adhere to all standards, resulting in problems in smart contracts. 
The ERC-20 standard is one of the APIs used to manipulate cryptocurrencies, including how to transfer tokens between addresses and access token data ~\citep{erc202022standard}. When transferring tokens, for example, \texttt{transfer()}, \texttt{transferFrom()}, and \texttt{approve()} will return a \texttt{boolean} value to indicate whether the function succeeded, and many smart contracts cause transfer mistakes since they do not verify the return value. 

\begin{lstlisting}[language=Solidity,
                    numbers=left,
                    captionpos=b,
                    % belowskip = -0.2em,
                    caption=Transaction Dependency Instance,
                    label = lst:Transaction_Dependency,
                    % title=Listing 3: Transaction dependency vulnerability instance
                    ] 
contract Wallet{
  address public owner;
  constructor() payable{
    owner = msg.sender; }
  function transfer(address to, uint amount) public{
    require(tx.origin == owner);
    (bool sent,) = to.call.value(amount);
    require(sent,"Failed to send Ether"); } }
contract Attack{
  address payable public owner;
  Wallet w;
  constructor(Wallet wal){
    w = Wallet(wal);
    owner = payable(msg.sender); }
  function attack() public{
    w.transfer(owner,address(w).balance); } }
\end{lstlisting} 
\vspace{-1ex}

\subsubsection{Transaction State Dependency}
Contracts should check the permissions of certain sensitive invocations that use the global variable \texttt{tx.origin}, which points to the address in the entire call stack where the transaction was originally sent~\citep{RN28}. Assume the \texttt{Wallet} contract in Listing \ref{lst:Transaction_Dependency} sends a transaction to the \texttt{Attack} contract, and then the \texttt{attack()} function invokes the \texttt{transfer()} function in the \texttt{Wallet} contract, at which point \texttt{tx.origin} meets the detection in line 6, making the success of the attack.

\subsubsection{Suicidal and Greedy Contracts}
Smart contracts usually include a provision enabling the owner to commit \texttt{suicide} if the contract is challenged. The SELFDESTRUCT operational code (opcode) in a suicidal contract can ignore all contract code logic, even the \texttt{fallback()} function~\citep{RN65}. However, attackers utilize this feature to corrupt the logic of some contracts, which leads to restrictions on all other operations that depend on the contracts. For example, the Parity wallet was attacked by a suicidal contract in 2017~\citep{RN66}, which resulted in a permanent lock of all cryptocurrencies that transferred to the wallet before the wallet maintainer fixed the vulnerability.

Similar to the suicidal contract, the greedy contract locks up the ether, but it is alive. Greedy contracts do not have instructions related to the withdraw and send~\citep{nikolic2018finding}, such as \texttt{send}, and \texttt{transfer}, so it locks all ethers and cannot withdraw. Therefore, making sure there are means to get ether out before transferring it to a contract~\citep{RN34}. 

\begin{lstlisting}[language=Solidity,
                    numbers=left,
                    captionpos=b,
                    % aboveskip=0em,
                    % belowskip = -0.7em,
                    caption=Reentrant Vulnerability Instance,
                    label = lst:Reentrant_Vulnerability,
                    %title=Listing 2: Reentrant vulnerability instance
                    ] 
function payOut(address recipient, uint amount) returns(bool){
  if(msg.sender != owner || msg.value>0 ||(payOwnerOnly && recipient != owner))
    throw;
  if(recipient.call.value(amount)()){
    payOut(recipient, amount);
    return true;
  }else{
    return false;
  } 
}
\end{lstlisting}
\vspace{-1ex}

\subsubsection{Reentrancy Vulnerability}
The concept of threads does not exist in Solidity, so it cannot execute more than two operations concurrently. This means that when a contract initiates a call via \texttt{call()}, it must wait for the completion of the call before making the next call. However, it would be attacked if the callee contracts change the global state during the waiting~\citep{RN26}. The DAO attack leverages the recursive invocations to make the system keep cycling until internal assets run out. It exits in line 4 of Listing \ref{lst:Reentrant_Vulnerability}~\citep{RN27}, where the original \texttt{recipient} continues executing \texttt{call.value()} after a successful transfer.

\subsubsection{Unchecked External Call}
The return value or the arguments of an external call can affect the states of the code. Many contracts do not check the return value, which leads to vulnerabilities.
The mode of logic used in this bug is similar to that of misuse ERC-20 standard. When a function calls code logic outside the contract, it is equivalent to the entire runtime in a black box. At this point, failure to check the return value of the external call may cause the logic of the contract to break. For example, when multiple functions are nested, and the external call does not check, the return value of the internal call in time can go wrong~\citep{RN34}. 

Smart contracts in the DeFi trade by using external call functions including \texttt{delegatecall()}, \texttt{call()}, \texttt{send()}. More crucially, a failed external call in these methods results in a transaction not being rolled back, which can cause logical effects.

\subsubsection{Strict Balance Equality}
Equations are commonly used in programs to make decisions concerning contract logic. When an attacker employs some methods, such as a suicide transfer ether, to alter the state of the variables utilized in the equation, rendering the judgments of the equation incorrect, the attack affects the logic of the code that follows the equation. For example, in Listing \ref{lst:Strict_Balance}, when the balance in the account is 1 ether, it passes the check in line 2. In line 3, the attacker transfers ether into the account, causing the judgment to fail, so the transfer in line 4 does not follow the normal logic. It is a loophole caused by not fully checking the judgment conditions of the equation.

\begin{lstlisting}[language=Solidity,
                    numbers=left,
                    captionpos=b,
                    % belowskip = -0.7ex,
                    caption=Strict Balance Equality Instance,
                    label = lst:Strict_Balance,
                    ] 
function receive(address a) payable{
  if(msg.value > 1 ether) throw;
  if(this.balance == 1 ether){
    a.send(1 ether);
  } 
}
\end{lstlisting}
\vspace{-1ex}

\subsection{Application Layer Vulnerabilities}
\label{sec:Dec_exchage_v}
The application layer visualizes the state in the chain and interacts directly with the user. In this paper, we focus on DAPPs in the financial domain. DeFi applications generally suffer from price manipulation attacks similar to traditional centralized financial applications. With the current development, the problems in the application layer could be divided into lending market imperfection, cryptocurrency instability, design imperfection, and abusive transaction exposure in Figure~\ref{fig:Class_Smartcontract_Application_vul}.

\subsubsection{Lending Market Imperfection}
When the prices in the market are out of balance, it will result in bad debts for one of the participants in the lending market. To get more loans, attackers can boost the cryptocurrency exchange rate on the oracle by modifying the real-time price-related status before the loan is made. For example, an attacker can gain a larger quantity of tokens by directly manipulating token prices in the asset pool or increasing the price of collateral before lending~\citep{RN55}, putting the borrower in danger of bad debt.

\subsubsection{Cryptocurrency Instability}
The large fluctuations of cryptocurrencies come from many reasons, one of which is the Pump-and-Dump. The instability can easily trigger liquidation procedures. Exchanges have chosen stablecoins, which are tied to the price of real-world money, as the pricing standard to minimize losses, but they still exist as a risk. For example, a 99.98 \%  plunge in May 2022 in the price of the luna coin, whose value is tied to a stablecoin called Terra, left the entire crypto market with over \$700 million in collateral liquidated~\citep{luna2022liquidation}.

\subsubsection{Design Imperfection}
The attackers utilize incorrectly configured functionality or specific convenience features of DeFi platform exchanges \citep{wang2021towards}. Flash loan is designed as risk-free loans to be a convenient improvement to the loan that needs to borrow the flash loan, exchange it for currency, and repay the loan in an atomic transaction. For example, attackers borrow the flash loan to receive collateral at a premium and make a profit in this atomic transaction~\citep{RN62}, which results in bad debts for the users who borrow money from attackers.

\subsubsection{Abusive Transaction Exposure}
Exchanges disclose all transactions as soon as feasible to ensure completely behavioral transparency because off-chain matching services are not automated. Unfortunately, exchanges can restrict access to select users and launch Denial of Service (DoS) or Decentralized Denial of Service (DDoS) attacks~\citep{RN63} to dominate the market, audit transactions, and even front-run orders.

\section{Analysis of Attack Events}
\label{sec:attack_events}
\begin{figure}[ht]
    \centering
    \includegraphics[width=0.45\textwidth]{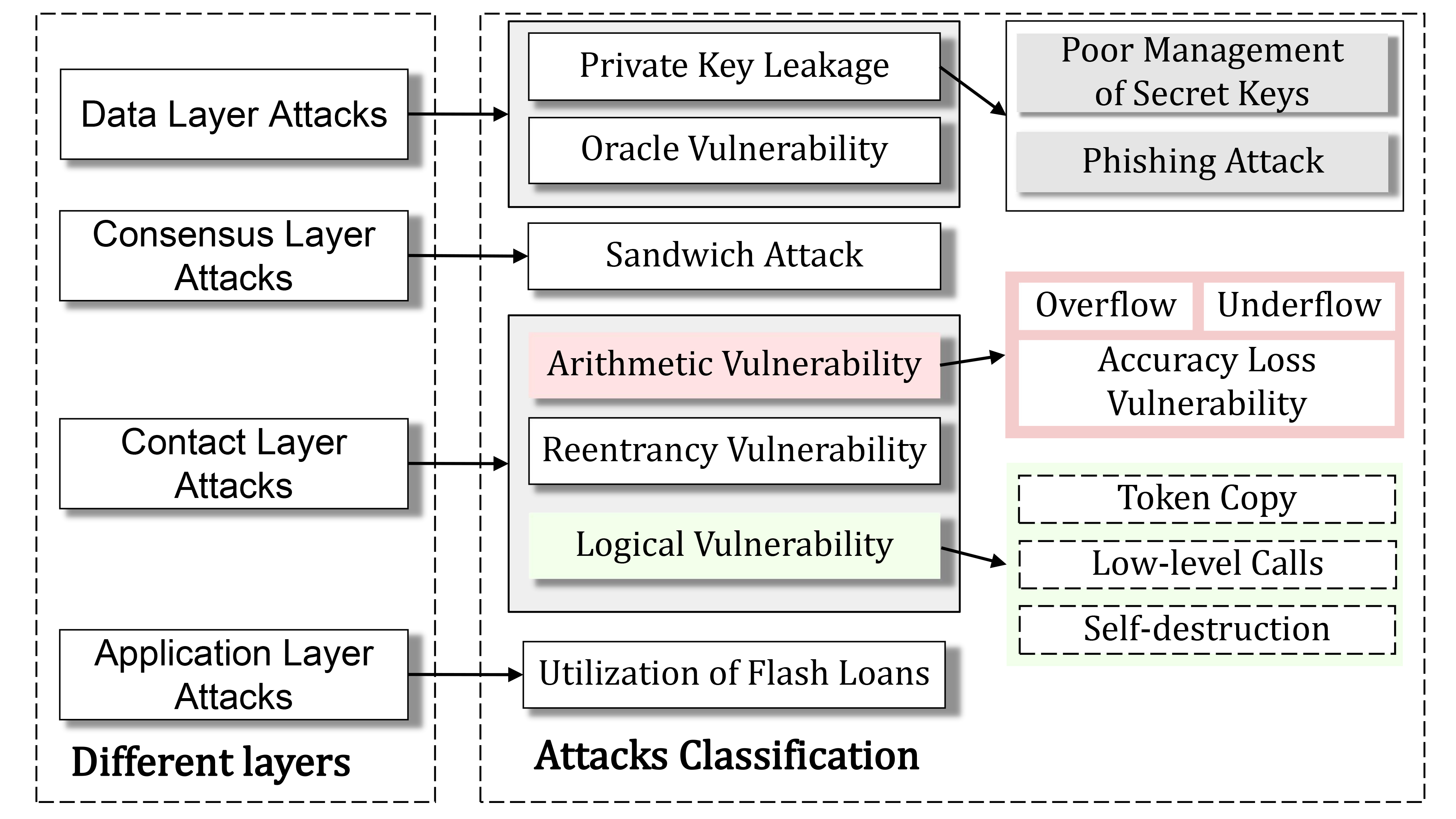}
    \caption{Classification of Attack Events at Each Layer}
    \vspace{-3ex}
    \label{fig:Attack_events}
\end{figure}

In this section, we investigate real-world attacks in the DeFi ecosystem~\citep{Documented-Timeline, RN8} and analyze the vulnerabilities exploited in the attacks with the classification shown in Figure~\ref{fig:Attack_events}.

\begin{table*}[width=18cm, pos=ht]
\setlength{\abovecaptionskip}{0cm} 
\setlength{\belowcaptionskip}{-0.2cm}
    \caption{Summarization Part of Real-world Attacks Exploiting Different Types of Vulnerabilities}
\footnotesize
\renewcommand{\arraystretch}{1.3}
\begin{tabular}{|p{2cm}| p{5cm} | p{2.4cm} |p{1.3cm} |p{2.1cm}| c|}
\ChangeRT{0.7pt}
\textbf{Vulnerabilities} & \textbf{Features}                                                       & \multicolumn{1}{l|}{\textbf{Victims}}                                                            & 
\multicolumn{1}{l|}{\textbf{Platform}}                                                          &
\multicolumn{1}{l|}{\textbf{Date}}                        
                      & \begin{tabular}[|c]{@{}c@{}}\textbf{Amount}\\ \textbf{(million USD)}\end{tabular}\\
\ChangeRT{0.5pt}
\multirow{12}{2.5cm}{Private Key Leakage} & 
        \multirow{12}{5cm}{The private keys of DeFi deployers are under threat due to poor private key management or phishing attacks. The key authorizes and verifies the transactions of the user. When an attacker utilizes the key, it is simple to tamper with the transaction, putting the trader's interests at risk. The attacker alters the website's Application Programming Interface (API) and embeds the vulnerability to get the user's personal information, including the user's key.}
        & Meerkat Finance& BSC      &March 2021     & 31\\
        & &Paid Network & Ethereum  &March 2021     & 160 \\
        & &Roll         & Ethereum  &March 2021     & 5 \\
        & &EasyFi       & Ploygon   &April 2021     & 80 \\
        & &bZx          & Ethereum  &November 2021  & 55 \\
        & &8ight Finance& Harmony   &December 2021  & 1 \\
        & &BitMart      & Ethereum  &December 2021  & 150\\
        & &AscendEX     & Ethereum  &December 2021  & 77 \\
        & &Vulcan Forged& BSC       &December 2021  & 140 \\
        & &LCX          & Ethereum  &January 2022   & 6 \\
        & &Ronin Bridge & Ethereum  &March 2022     & 624\\
        & &Poly Network & BSC       &September 2021 & 600\\
        
\ChangeRT{0.5pt}    
\multirow{6}{2.5cm}{Oracle Attacks}   & 
        \multirow{6}{5cm}{The oracle price data feed can be manipulated by the attackers who change the asset data for the smart contracts. When an oracle is attacked, real-world data posted to the blockchain changes. It mismatches on-chain data with the real world, harming users.}            
        &bZx  & Ethereum    & February 2020  & 0.9\\
        & &Harvest Finance  & BSC       &October 2020   &24\\
        & &Cheese Bank      & Ethereum  &November 2020  &3\\
        & &PancakeBunny     & BSC       &July 2021      &2\\
        & &Vee Finance      & Multichain&September 2021 &35\\
        & &Vesper Finance   & Ethereum  &December 2021  &1\\
\ChangeRT{0.5pt}
\multirow{2}{2cm}{DDoS Attacks}          & 
        \multirow{2}{5cm}{The attacker uses massive throughput to disrupt the device or server.}
        & McAfee DEX    & Multichain    & October 2019      & --\\
        & &  Youbi DEX  & Multichain    & November 2020     & --\\
\ChangeRT{0.5pt}
\multirow{2}{2cm}{Sandwich Attacks}          & 
        \multirow{2}{5cm}{Attackers use two transactions to clip victim's transaction and profit from it.}
        & Uniswap  & Ethereum   & May 2021 & 0.2\\
        & &  Uniswap \& Linch   & Ethereum & August 2022 & 3.7\\
\ChangeRT{0.5pt}
\multirow{3}{2cm}{51\% Attacks}          & 
        \multirow{3}{5cm}{Attackers can create fraudulently some transactions, when they control over 50\% of the blockchain's computing power.}
        & Ethereum Classic  & Ethereum  &January 2019   & 5.6\\
        
        & &  ZenCash        & Zendoo    & June 2018     & 0.5\\
        & &  PegNet         & Multichain&April 2020     & 0.6\\
\ChangeRT{0.5pt}
\multirow{2}{2cm}{Sybil Attack}              &
        \multirow{2}{5cm}{Attackers create multiple fake nodes to influence the blockchain network state.}
        & Solana  & Solana   &  August 2022 & --\\
        & &Ribbon Finance & Ethereum & October 2021& 2.5 \\

\ChangeRT{0.7pt} 
\end{tabular}
\vspace{-1ex}
  \label{tab:realworld_attacks1}
\end{table*}

\subsection{Date Layer Events}
\setParDis
\subsubsection{Utilization of Private Key Leakage}
\setParDef
The developer deploys DeFi applications on blockchain through private keys managed in the wallet. Also, users confirm and initiate transactions on the DeFi app through the private key. We summarize real-world DeFi security events due to private key leaks in Table \ref{tab:realworld_attacks1}. We believe there are two reasons for these security incidents: (1) poor management of secret keys; (2) phishing attacks.

\textbf{Poor Management of Secret Keys.} In the Meerkat Finance~\citep{Meerkat} incident, the administrator of the project used a private key and a false time lock in the contract. It transferred about \$30 million worth of \uppercase{BNB} tokens from the \uppercase{BNB} Vault. In Listing \ref{lst:Meerkat_Finance}~\citep{MeerkatCode}, the administrator used the visual ambiguity of the number "0" and the letter "o" to make the variable slot values in the \texttt{\underline{ }admin()} and \texttt{\underline{ }setAdmin()} functions differently. This means that the time lock of \uppercase{BNB} Vault is false, and the administrator can transfer the \uppercase{BNB} tokens via the backdoor.

\begin{lstlisting}[language=Solidity,
                    numbers=left,
                    captionpos=b,
                    % aboveskip = 0.5em,
                    belowskip = 2em,
                    caption=\textsc{Meerkat Finance} Instance,
                    label = lst:Meerkat_Finance
                    %title=Listing 2: Reentrant vulnerability instance
                    ] 
function _admin() internal view returns (address adm) {
    bytes32 slot = ADMIN_SLOT;
    assembly { 
      adm := sload(slot) } 
}
function _setAdmin(address newAdmin) internal {
    bytes32 slot = ADMIN_SL0T;
    assembly { 
      sstore(slot, newAdmin) } 
}
\end{lstlisting}
\vspace{-2ex}

\textbf{Phishing Attack.} The scripts embedded in the DeFi website interact with the wallet, which may give opportunities for phishing attacks~\citep{winter2021s}. In the BadgerDAO incident~\citep{badger}, the attackers stole the Badger developer's secret keys and injected malicious scripts into BadgerDAO's web pages. The scripts intercepted the user's transactions and prompted the user to allow the attacker to operate on the ERC-20 tokens in their wallets.

The transparent nature of DeFi allowed the attacker to easily gather information about the developers. The attacker sent malicious emails to bZx developers, stealing the private management key of bZx deployed on the BSC and Polygon chains. The attackers used the management private key to upgrade the contract to mint unlimited tokens~\citep{bZx}.

\textbf{Backdoor Attack.}
A backdoor attack involves unauthorized access to a program or system that bypasses the software's security checks. The input is only sent to the attacker's subtask when the set trigger is activated. This attack does not affect the system's regular operation, so only the attacker can trigger and profit from it.
Furthermore, systems that automatically collect and process data, such as oracle, can reach the problem at the functional level. It can be through the introduction of data with hidden vulnerabilities in large quantities, in addition to the attacker inserting his event at the code level.

The Ronin Bridge network incident in Table \ref{tab:realworld_attacks1}~\citep{RN23}, which is a sidechain of Ethereum, was attacked by a backdoor attack in March 2022. It established a new record for the most significant losses in the DeFi space with 624 million USD. There are nine authentication nodes in the Ronin chain, with access being granted after getting verification at five of them. Through a backdoor, the attacker has access to and control over the five private keys, the authentication, and thus the withdrawal event.

\textbf{Hash Collision.}
When compiling the smart contract source code into bytecode, the first 4 bytes of the hash of the method name are used as the token. It can be utilized by an attacker to generate a signature that satisfies the specified 4-bytes token through a hash collision. When a contract can be executed by passing method names as parameters, it is possible to hack the contract by using the hash collision. The poly network case is where the attacker constructs specific method signatures through hash collisions to execute some special functions as a contract manager~\citep{poly-network}.

In the PloyNetwork event~\citep{poly-network} on the BSC chain, the attackers used hash collisions to construct Keeper's signatures and modified Keeper's public keys through the \texttt{putCurEpochConPubKeyBytes()} function in the management contract. This incident resulted in a loss of 600 million USD.

\subsubsection{Utilization of Oracle Vulnerability}
The DeFi ecosystem relies heavily on oracle to provide off-chain or on-chain asset data, and cannot verify the accuracy of the data. This means that if the DeFi protocol uses only a single DEX as the source of asset prices, then the DeFi protocol will assume that it is true and accurate regardless of the movement of its asset price data.

In Table \ref{tab:realworld_attacks1}, oracle attacks have caused significant damage to DeFi applications. Most of the oracle attacks are based on the following steps~\citep{wang2021promutator}.

\begin{itemize}
\item [(i)] \textbf{Preparation of Funds.} The attacker borrows a large number of assets unsecured through various Flash Loan providers, e.g., bZx, dYdX. He/She intends to inject the assets into other DeFi agreements to inflate their prices while hoarding the target assets.

\item [(ii)] \textbf{Raising the Price of Target Assets.} The attacker manipulates the oracle by balancing the target assets stored in the liquidity pool, i.e., by exchanging a large number of tokens back and forth between different liquidity pools. Since a single oracle is used, it passes the manipulated price data into the DeFi protocol.

\item [(iii)] \textbf{Profiting.} The attacker exchanges the target asset for money borrowed by the Flash Loan, a service provided by DeFi, e.g., collateralized borrowing. As the attacker inflates the price of the target asset, it can exchange the target asset for a larger amount of other assets. By this step, the attacker will gain much profit.

\item [(iv)] \textbf{Loan Repayment.} The attacker restores the assets in the liquidity pool to their initial state to avoid losses caused by price slippage~\citep{wang2021promutator}, and repays the loan.
\end{itemize}

The bZx attack~\citep{bZxattackanalysis} happened in February 2020, and it was through the above attack steps that the attackers made a profit of about \$0.9 million. The attacker borrowed lots of ETH through the bZx platform. At KyberSwap AMM, a portion of the ETH was exchanged for sUSD tokens to drive up the price of sUSD. Next, the attacker bought the sUSD from the Synthetic Depot contract at the normal price. The attacker pledged the sUSD in the account into the bZx protocol in exchange for ETH. As the price of sUSD in bZx was inflated, it could be exchanged for more ETH. Finally, the attacker repaid the loan.

In 2021, Vee Finance lost 35 million USD due to the oracle vulnerability~\citep{RN47}. It had only one oracle as a price input source. At the same time, the attackers profited by using errors in the contract to bypass the slippage protection checks. Similarly, the Harvest protocol used the USDT price in Curve as the price data. Since the USDT price became lower at this point, the attacker could pledge more USDT with the same assets. The attacker performed 32 attacks and profited 24 million USD from the protocol.

\subsubsection{Distributed Denial of Service Attack Events}
\label{subsec:DDoS}
A Distributed Denial of Service (DDoS) attack can be a DOS attack performed on multiple nodes to some extent. Due to the large number of nodes and the unpredictability of block-generating nodes, DoS attacks on blockchain systems such as Ethereum are less detrimental. However, the DDoS attack can typically cause network nodes to fail or slow down transactions across the network by running out of memory due to many transaction requests. It causes these nodes to be unable to close the transaction they are processing, and other nodes are unable to perform the transaction. Attackers typically combine other attacks to cause damage to blockchain-based DeFi applications.

Numerous DeFi applications in Table \ref{tab:realworld_attacks1}, such as McAfee \citep{DDOS2021} and Youbi DEX~\citep{Yobi2020}, have recently been subjected to DDoS attacks, causing substantial harm to the DeFi ecosystem. At the initial launch, the McAfee DEX was under a DDoS attack. Due to the number of user nodes being insufficient, the failure of some users had a more significant impact on the entire network.

\subsection{Consensus Layer Events}
\setParDis
\subsubsection{Sandwich Attack Events}
\setParDef
\label{subsec:sandwich_attack}
Sandwich attack capitalizes on miners' pursuit of MEV by reordering transactions to achieve the attack's objective. According to the consensus rule, the node with control over the block can choose the order of transactions. The dependency on transaction order vulnerability is one of the factors affecting the security of smart contracts, and it also applies to DeFi applications.

The Sandwich attack applies to AMMs like Uniswap and takes advantage of a special feature of AMMs, such as the fact that for every token swap that occurs on an AMM like Uniswap, the price of its swapped tokens changes. The steps of the Sandwich attack are as follows:

\begin{itemize}
\item [(i)] \textbf{Network Spy.} There are some spy nodes deployed on the network to collect all the transactions for asset exchanges. If attackers consider that a transaction that exchanges token $A$ for token $B$ is profitable, they will create two transactions for racing to control the transaction and make a profit. It means that the price of token $B$ in the liquidity pool will be increased.

\item [(ii)] \textbf{Transaction Creation.} The attacker creates a front-running transaction to exchange token $A$ for token $B$, and the price of token $B$ in the liquidity pool will be raised. Suppose the price of token $B$ rises too much. In that case, the slippage detection may be triggered, and the attack will be failed, so the attacker will generally control the number of tokens purchased. The victim is also exchanging token $A$ for token $B$, which causes the price of token $B$ to continue to rise. As the attacker's front-running trading causes the price of token $B$ to rise, the victim can only obtain less than the expected amount of token $B$. Finally, after the victim's transaction, the attackers would create a back-running transaction that converts token $B$ into token $A$, thus making a profit.
\end{itemize}

According to our research, sandwich attacks often occur on AMMs, such as Uniswap, Linch, and SushiSwap. About 4 thousand sandwich attacks have occurred on Ethereum, which allowed the attackers to generate a profit of 3.7 million USD~\citep{sandwichattack}.

\subsubsection{51\% Attack Events}
To achieve the 51\% attack, the attacker creates a transaction that transfers tokens to the victim~\citep{51-attack}. Concurrently, the attacker generates an alternative chain. If more than 51\% of nodes in the network support the alternative chain generated by the attacker, the consensus mechanism deems the original chain invalid. The attack requires most of the computing power in the whole network, so the greater the computing power of honest nodes on blockchain platforms, such as Bitcoin, the more difficult it is to profit from the attack~\citep{Crypto512022}.

In Table \ref{tab:realworld_attacks1}, we gathered some DeFi projects with 51\% attacks. Considering the significance of the situation, the amount of damage is not particularly severe. The 51\% attack exploits a well-known vulnerability that manifests itself frequently in decentralized exchanges.

\subsubsection{Sybil Attack Events}
To execute a Sybil attack, the attacker typically chooses to control the network through multiple accounts or nodes. The attack slows down the entire network and can impact DeFi transactions when the number of affected nodes exceeds a certain threshold. 

In Table \ref{tab:realworld_attacks1}, a Sybil attack in 2022 was observed on Solana, where the attacker kept constructing a network of protocols to deceive asset owners into believing they were popular, resulting in the TVL of 7.5 billion USD.
Ribbon Finance is an option-based financial program. It is constantly creating as many wallets as possible in 2021 to profit from airdrops to users with more than \$100. The Sybil attack has significantly impacted asset market users in terms of their regular trading and losses.

\subsection{Contract Layer Events}
Smart contracts are the basis for decentralized financial instruments. When the DeFi applications were deployed on the blockchain~\citep{torres2018osiris}, it was possible that smart contract errors would cause irreparable harm to DeFi.

\subsubsection{Utilization of Arithmetic Vulnerability}
Almost every DeFi use case involves performing arithmetic operations on different forms of currency. Among these operations are adjusting account balances by adding or subtracting amounts and converting exchange rates between various tokens~\citep{RN47}. In the DeFi ecosystem, overflow and precision loss vulnerabilities have been identified. These arithmetic bugs have caused significant damage.

\textbf{Overflow.} In April 2018, multiple DeFi applications on Ethereum, including OKEx, were forced to shut down because of an overflow vulnerability in an ERC-20 token contract. MESH and UGToken were among these applications. OKEx was one of the applications forced to close due to astronomical losses. This overflow event shares several characteristics, including the difficulty caused by the \texttt{transferProxy()} function in Listing \ref{lst:Snippets_of_MESH}~\citep{MESH}.
\begin{lstlisting}[language=Solidity,
                    numbers=left,
                    captionpos=b,
                    % aboveskip = 0.5em,
                    belowskip = 2em,
                    caption=Snippets of \textsc{MESH Token} Contract,
                    label = lst:Snippets_of_MESH
                    ] 
function transferProxy(address _from, address _to, uint256 _value, uint256 _fee,uint8 _v,bytes32 _r, bytes32 _s) public transferAllowed(_from) returns (bool){
  if(balances[_from] < _fee + _value) revert();
  ...
  Transfer(_from, _to, _value);
  balances[msg.sender] += _fee;
  Transfer(_from, msg.sender, _fee);
  ...
  return true;
}
\end{lstlisting}
\vspace{-1ex}

In line 2 of Listing \ref{lst:Snippets_of_MESH}~\citep{MESH}, it contains the potentially dangerous overflow vulnerability. Since both \texttt{\underline{ }fee} and \texttt{\underline{ }value} are input parameters, and they are susceptible to be manipulated by humans. An adversary can then construct the incoming parameters so that their size exceeds the storage range of the \texttt{uint} data type, resulting in an overflow. When an overflow occurs, the entire value of the unsigned integer becomes 0. It indicates that an attacker can bypass the check performed by the \texttt{if} statement on the second line~\citep{transferProxy} and cause tokens to be transferred to an empty address.

\textbf{Underflow.} The larger loss in arithmetic vulnerability is Compound Finance. Its reward payouts \texttt{CompSpeed} could be set to 0, which indicated that reward payouts were suspended, and the market award index \texttt{supplyIndex} was 0. For new users, their award index \texttt{supplierIndex} was initialized to \texttt{CompInitialIndex} presented by Compound as $10^{36}$. An underflow vulnerability occurred in Listing \ref{lst:Snippets_of_Compound}~\citep{compound} at line 8. This caused the formula for calculating the difference in the index \texttt{deltaIndex = sub\_(supplierIndex = 0, supplierIndex=$10^{36}$)} to underflow and became a very large value, while the Compound Finance reward calculation relied on the value of \texttt{deltaIndex}.

\begin{lstlisting}[language=Solidity,
                    numbers=left,
                    captionpos=b,
                    % aboveskip = 1em,
                    belowskip = 2em,
                    caption=Snippets of \textsc{Compound} Contract,
                    label = lst:Snippets_of_Compound
                    ] 
function _setCompSpeed(CToken cToken, uint compSpeed) public {
  ...
  setCompSpeedInternal(cToken, compSpeed);
}
  ...
  if (supplierIndex.mantissa == 0 && supplyIndex.mantissa > 0) {
    ... }
  Double memory deltaIndex = sub_(supplyIndex, supplierIndex);
\end{lstlisting}
\vspace{-1ex}

There was no attacker in this security incident, but rather an overpayment of rewards due to an underflow vulnerability in the contract. 
This incident caused the Compound 80 million USD in damages. In 2022, Umbrella NetWork also lost 0.7 million USD due to an underflow vulnerability.

\begin{table*}[width=18cm, pos=ht]
\setlength{\abovecaptionskip}{0cm} 
    \caption{Summarization Part of Real-world Attacks Exploiting Different Types of Vulnerabilities}
\footnotesize
\renewcommand{\arraystretch}{1.3}
\begin{tabular}{|p{2cm}| p{5cm} | p{2.4cm} |p{1.3cm} |p{2.1cm}| c|}
\ChangeRT{0.7pt}
\textbf{Vulnerabilities} & \textbf{Features}                                                       & \multicolumn{1}{l|}{\textbf{Victims}}                                                            & 
\multicolumn{1}{l|}{\textbf{Platform}}                                                          &
\multicolumn{1}{l|}{\textbf{Date}}                        
                      & \begin{tabular}[|c]{@{}c@{}}\textbf{Amount}\\ \textbf{(million USD)}\end{tabular}\\
\ChangeRT{0.5pt}
\multirow{4}{2cm}{Arithmetic Vulnerability}    & 
        \multirow{4}{5cm}{The attacker passes in specific parameters that cause the arithmetic operations in the contract to overflow. }
        &Uranium Finance    & BNB   &April 2020 & 50\\ 
        & &Compound         & Ethereum      &September 2021 & 80 \\
        & &Pizza DeFi       & EOS           &December 2021  & 5 \\
        & &Umbrella Network & Ethereum      &March 2022     & 0.7  \\
\ChangeRT{0.5pt}
\multirow{3}{3cm}{Reentrancy Vulnerability}     & 
         \multirow{3}{5cm}{When a function calls an untrusted contract and that contract recursively calls the original function, it's reentrant.} 
        & dForce & Ethereum&April 2020 &24 \\
        & &Akropolis    & Ethereum  & November 2020 &2\\
        & &Grim Finance & Fantom    &December 2021  &30\\
\ChangeRT{0.5pt}
\multirow{9}{2cm}{Logical Vulnerability}& 
        \multirow{9}{5cm}{The adversary employs unique methods to alter the contract program logic inadvertently and cause the loss of the DeFi application. It comprises possessing a token copy, low-level calls, self-destroying, and transaction rollback attack.}       
        &  Betdice              & EOS       &December 2018  &3\\
        & &EOS MAX              & EOS       &December 2018  &0.9\\
        & &Ethereum Classic     & Ethereum  &January 2019   &0.5\\
        & &Furucombo            & Ethereum  &February 2020  &14\\
        & &bZx                  & Ethereum  &November 2020  &8\\
        & &BurgerSwap           & BNB       &May 2021       &7\\
        & &Eleven Finance       & Polygon   &June 2021      &4\\
        & &Punk Protocol        & Ethereum  &August 2021    &3\\
        & &Starstream Finance   & Ethereum  &April 2022     &4\\

\ChangeRT{0.5pt}
\multirow{3}{2cm}{Flash Loan}          & 
        \multirow{3}{5cm}{It allows users to borrow and settle loans in real-time in a single transaction without providing any collateral.}
        & Warp Finance      & Ethereum  &December 2020  &7\\
        & & Alpha Homora    & Ethereum  &February 2021  &37\\
        & & Elephant Money  & BNB       & April 2022    &11\\
        
\ChangeRT{0.5pt}
\multirow{3}{2cm}{Pump-and-Dump attack}          & 
        \multirow{3}{5cm}{The attacker organizes many individuals torise significantly for a brief period and profits from it.}
        & Cryptopia DEX  & Multichain  & November 2018  & 157\\
        & & Binance DEX  & BNB  & May 2019  &50\\
        & & Squid        & BNB  & November 2021  &300 \\
\ChangeRT{0.7pt} 
\end{tabular}
  \label{tab:realworld_attacks2}
\end{table*}

\textbf{Accuracy Loss Vulnerability.} The Uranium Finance contract allowed users to borrow money using Flash Loan. However, the contract suffered from accuracy handling errors when calculating the amount to be returned, resulting in the amount that was 100 times larger than expectation \citep{uranium-finances-hacked}. The attacker only needs to return a small portion of the loan to pass the check of the \texttt{require} statement in Listing \ref{lst:Snippets_of_UraniumPair}~\citep{UraniumPair} and pays off the loan.

\begin{lstlisting}[language=Solidity,
                    numbers=left,
                    captionpos=b,
                    % aboveskip = 0.5em,
                    % belowskip = 1.5em,
                    caption=Snippets of \textsc{UraniumPair} Contract,
                    label = lst:Snippets_of_UraniumPair
                    ] 
uint balance0Adjusted = balance0.mul(10000).sub(amount0In.mul(16));
uint balance1Adjusted = balance1.mul(10000).sub(amount1In.mul(16));
require(balance0Adjusted.mul(balance1Adjusted) >=uint(_reserve0).mul(_reserve1).mul(1000**2), 'UraniumSwap: K');
\end{lstlisting}

\subsubsection{Utilization of Reentrancy Vulnerability}
A contract executing a transaction invokes a malicious contract account, and the malicious contract account invokes a function in the contract before the contract state changes \citep{RN93}.
 The most significant reentrancy attack on Ethereum was the DAO attack~\citep{RN24}, which caused a hard fork of Ethereum by constantly calling the \texttt{withdrawBalance()} function to achieve an infinite transfer operation. Reentrancy attacks were applied to the DeFi protocol with its development.
In Table \ref{tab:realworld_attacks2}, 54 million USD was lost to DeFi due to a reentrancy vulnerability.

In April 2020, the dForce protocol suffered a reentrancy, losing about 24 million USD. The attackers exploited the ERC-777~\citep{erc-7772022standard} compliant imBTC tokens. Compared to the ERC-20 token standard, the ERC-777 token standard has one feature. When ERC-777 tokens were sent or received, they would go through Hook in the form of a callback to notify the sender or recipient. The attacker in the incident~\citep{RN47} took advantage of this feature and re-entered the dForce contract to increase the amount of imBTC collateral and get a higher yield.

Grim Finance on the Fantom~\citep{fantom} chain lost 30 million USD due to a re-entry vulnerability. First, the attacker created a contract to inject the cryptocurrency borrowed through the Flash Loan service into Spirit Swap~\citep{SpiritSwap} to obtain Spirit-LP certificates. Next, the Spirit-LP certificates were pledged to the GrimBoostVault contract in exchange for the GB-BTC-FTM, which was a token, via the \texttt{depositFor()} function in Listing \ref{lst:Snippets_of_GrimBoostVault}~\citep{GrimBoost}. Since the legitimacy of the token contract was not verified, the attacker re-called the \texttt{depositFor()} function in the \texttt{safeTransferFrom()} function of the malicious contract, implementing reentrancy to collateralize more GB-BTC-FTM for profit. The attacker has finally paid back the loan.

\begin{lstlisting}[language=Solidity,
                    numbers=left,
                    captionpos=b,
                    belowskip= 1em,
                    caption=Snippets of \textsc{GrimBoostVault} Contract,
                    label = lst:Snippets_of_GrimBoostVault
                    ] 
function depositFor(address token, uint _amount, address user) public {
  uint256 _pool = balance();
  IERC20(token).safeTransferFrom(msg.sender, address(this), _amount);
  earn();
  uint256 _after = balance();
  _amount = _after.sub(_pool); 
  uint256 shares = 0;
  if (totalSupply() == 0) {
    shares = _amount;} 
  else{ shares = (_amount.mul(totalSupply())).div(_pool); }
  _mint(user, shares); } 
}
\end{lstlisting}
\vspace{-2ex}

\subsubsection{Utilization of Logical Vulnerability}
According to our investigation, a large number of vulnerabilities in the DeFi application stem from the simple programming errors in the smart contract~\citep{RN47}. Due to the tamper-evident nature of the blockchain, these errors can cause significant damage to the DeFi application.

\textbf{Token Copy.} This was the third attack on bZx in 2020. The attackers exploited a vulnerability in the contract by passing the same address to the sender parameter \texttt{\underline{ }balancesFrom} and the receiver parameter \texttt{\underline{ }balancesTo} in the bZx contract, thus copying the balance in the account~\citep{iToken}.

\textbf{Low-level Calls.} Starstream Finance on Ethereum is a DeFi project on the Metis Andromeda network. As seen in Listing \ref{lst:Snippets_of_DistributorTreasury}~\citep{DistributorTreasury}, the vulnerability was due to the public function \texttt{execute()} of the DistributorTreasury contract using an unchecked external call \texttt{to.call()}, allowing anyone to make an external call. It meant that an attacker could use the function to generate a call to the \texttt{withdrawTokens()} function to extract the STAR Token in the StarstreamTreasury contract.

\begin{lstlisting}[language=Solidity,
                    numbers=left,
                    captionpos=b,
                    % aboveskip = 0.5em,
                    % belowskip = 1em,
                    caption=Snippets of \textsc{DistributorTreasury} Contract,
                    label = lst:Snippets_of_DistributorTreasury
                    ] 
function execute(address to,uint256 value,bytes calldata data) external returns (bool, bytes memory) {
  (bool success, bytes memory result) = to.call{value: value}(data);
    return (success, result);
}
\end{lstlisting}

\textbf{Self-destruction.} Self-destruction of contracts and destruction of tokens in contracts are both common operations in the DeFi ecosystem. Usually, attackers will transfer stolen valuable cryptocurrency into the contract under their control. To avoid being traced, the attackers will destroy the attack contract after transferring the tokens in their contracts.

\begin{lstlisting}[language=Solidity,
                    numbers=left,
                    captionpos=b,
                    % aboveskip = 0.5em,
                    % belowskip = 2em,
                    caption=Snippets of \textsc{NeverSellVault} Contract,
                    label = lst:Snippets_of_ElevenNeverSellVault
                    ] 
function withdraw(uint256 _shares) public {
  ...
  if(avai<_shares) IMasterMind(mastermind).withdraw(nrvPid, (_shares.sub(avai)));
    token.safeTransfer(msg.sender, _shares);
    ...}
function emergencyBurn() public {
  ...
  if(avai<balan)
    IMasterMind(mastermind).withdraw(nrvPid, (balan.sub(avai)));
  token.safeTransfer(msg.sender, balan);
  ...}
\end{lstlisting}

The Eleven Finance attack~\citep{ElevenFinance} was caused by the fact that the attacker could not destroy the proof of assets when withdrawing them from the contract, thus enabling the withdrawal of the deposit twice. The specific reason for this attack was that the \texttt{emergencyBurn()} function in the ElevenNeverSellV insurance contract allowed the attacker to withdraw the deposited assets without destroying their proofs. Afterward, the attacker called the \texttt{withdraw()} function in Listing \ref{lst:Snippets_of_ElevenNeverSellVault}~\citep{ElevenNeverSellVault}  to perform the normal process of withdrawing the assets. This incident caused a loss of approximately 4 million USD to Eleven Finance on the Polygon chain.

\textbf{Transaction Rollback Attack.} Several blockchain platforms, including TORN, EOS, etc., have been subject to transaction rollback attacks. Hackers use an inline function to undo a transaction. There are multiple activities within a transaction; as long as one activity is abnormal, all the activities in the transaction will fail. For example, in 2019, attackers in Ethereum Classic rolled back transactions by coding their contract service~\citep{TransactionRollBack2019}. They continued attempting until the result of the function call satisfied the requirements. As the result in Table \ref{tab:realworld_attacks2} shows, multiple double spending vulnerabilities were introduced into Ethereum Classic, acquiring approximately 0.5 million USD.

\subsection{Application Layer Events}
Some application layer attacks are caused by the lending market imperfections and abusive transaction exposure, whereas design imperfections and cryptocurrency instability directly cause others. The Pump-and-dump attack is a result of the cryptocurrency instability, and the use of Flash Loan is a result of the design imperfections. Some DDoS attacks in Section \ref{subsec:DDoS} cause the lending market imperfections, and the sandwich attacks caused by the abusive transaction exposure are in Section \ref{subsec:sandwich_attack}.

\subsubsection{Utilization of Flash Loans}
Flash Loans are an unsecured form of lending that adds vitality to DeFi. Blockchain transactions' atomicity validates these loans' legitimacy at the execution time~\citep{qin2021attacking}. Unfortunately, attackers have access to Flash Loan, reducing the cost of their attack. According to our survey results, most DeFi attacks targeted Flash Loan services.

At the same time, Flash Loan protects the assets of arbitrageurs so they can use DeFi to manipulate prices on traditional financial markets. The arbitrageur employs a portion of the borrowed assets to increase or decrease the price of the assets included in the AMM liquidity pool, also known as the asset exchange ratio. According to~\citep{RN55}, the arbitrageurs then use the remaining loan funds to engage in another profitable trade. They have ultimately repaid the loan. In April 2022, the Elephant Money attack was a classic example of price manipulation, resulting in 11 million USD lost. The arbitrageurs increased the supply of the TRUNK stablecoin, causing the ELEPHANT token's price to rise~\citep{ElephantMoney}. Ultimately, they made a profit by purchasing additional WBNB and BUSD tokens with ELEPHANT and TRUNK.

\subsubsection{Pump-and-Dump Attack Events}
There have been some trading attacks on the cryptocurrency market that has caused the price to fluctuate abnormally, one of which is the Pump-and-Dump attack~\citep{xu2019anatomy,kamps2018moon}. Pump-and-Dump attack is comprised of the following five stages.
(i) The organizer initially organizes a group of individuals to prevent them from engaging in the exchange. (ii) Moreover, to increase the revenue, the organizer aggressively promotes the purchase of the target so that the majority of people buy cryptocurrencies at the same time at time $t$, instilling confidence in the buyers. (iii) When the time $t$ arrives, everyone begins to purchase the target, causing the price to rise sharply. Furthermore, the price typically reaches its peak within minutes. (iv) After a price increase, the price typically falls lower than the initial price, and some unwilling sellers become trapped. (v) After the event, the organizers will boast in inaccessible locations about how much the event's peak was raised. It misleads newcomers into believing that it can be profitable.

Four of the most common exchanges for Pump-and-Dump attacks are Binance, Bittrex, Cryptopia, and Yobit~\citep{xu2019anatomy}. In Table \ref{tab:realworld_attacks2}, we have collected three real-world Pump and Dump attacks containing Cryptopia DEX, Binance DEX, and Squid. For instance, Cryptopia~\citep{xu2019anatomy} advertised on Telegram, a widespread social media platform, that it would Pump a particular asset at a specific moment and begin a countdown. This action attracted a large number of Profit-seekers simultaneously, causing the value of the assets to increase. However, after reaching the peak, the organizers began arbitraging. Since the victims had faith in the asset's upward trend, they lacked a stop loss operation when the price dropped. In reality, the victims remain at risk and are entrapped.

\section{Analysis of Security Optimizations}
\label{Security_Optimization}
\begin{figure}[htbp]
\setlength{\abovecaptionskip}{0cm} 
\setlength{\belowcaptionskip}{-2cm}
    \centering
    \vspace{1ex}
    \includegraphics[width=6.5cm]{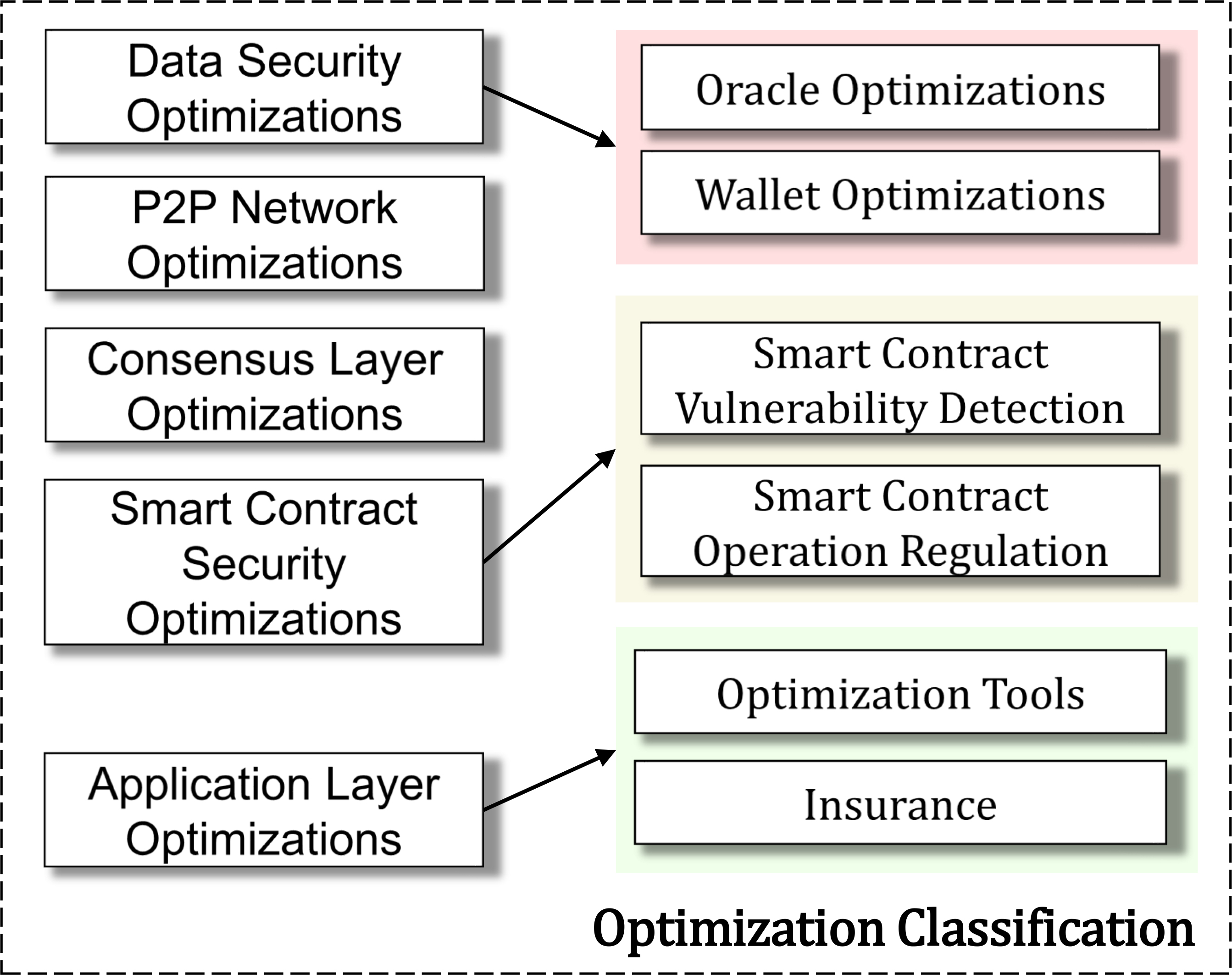}
    \caption{Classification of Optimization at Each Layer}
    \vspace{-3ex}
    \label{fig:Class_Optimization}
\end{figure}

Even though numerous attacks exploit various vulnerabilities, many research efforts have succeeded in detecting and defending against these attacks, which fuels the rapid development of blockchain. We analyze the vulnerabilities in the Section \ref{sec:ana_v} according to the hierarchy shown in Figure~\ref{fig:Class_Optimization}.

\subsection{Data Security Optimizations}
As the oracle mechanism and key management flaws are mentioned in Section \ref{sec:data_security_v}, optimizing the data layer can effectively prevent attacks from exploiting the flaws to tamper with the authenticity of the data.

\subsubsection{Oracle Optimization Schemes}

\begin{table*}[width=18cm, pos=htbp]
\setlength{\abovecaptionskip}{0cm} 
\setlength{\belowcaptionskip}{-0.2cm}
\centering
\caption{Summarization of Different Oracle Optimization Schemes.}
\footnotesize
\renewcommand{\arraystretch}{1.3} 
\begin{tabular}{|p{2.2cm}| p{2cm}| p{6cm}| c| c| c| c| c|} 
\ChangeRT{0.7pt}
      \multirow{2}{3cm}{\textbf{Reference}} & \multirow{2}{2cm}{\textbf{Key Technologies}} & \multirow{2}{*}{\textbf{Features}} & \multicolumn{5}{c|}{\textbf{Categories}}   \\
      \cline{4-8}
      &  & & C-On & C-Off & D-On & D-Off & Others \\
\ChangeRT{0.5pt}
  \multirow{2}{2cm}{\cite{zhang2016town}} &\multirow{2}{2cm}{TEE} & \multirow{2}{5.5cm}{Operations are handled in TEE to invalidate malicious requests at cost of throughput}& \multirow{2}{*}{\CheckmarkBold} &\multirow{2}{*}{$\times$} &\multirow{2}{*}{$\times$}  &\multirow{2}{*}{$\times$}  &\multirow{2}{*}{$\times$}  \\
  &  & & & & & &\\
\ChangeRT{0.05pt}
  
  \multirow{2}{3cm}{\cite{chainlink2022}}& \multirow{2}{2cm}{Node Network} & \multirow{2}{5.5cm}{The median value of multi-parties stored in the node network to prevent tampering}  &\multirow{2}{*}{$\times$} &\multirow{2}{*}{$\times$} & \multirow{2}{*}{\CheckmarkBold} &\multirow{2}{*}{$\times$} &\multirow{2}{*}{$\times$} \\
  &  & & & & & &\\ 
\ChangeRT{0.05pt}
  
  \multirow{2}{2cm}{\cite{ritzdorf2017tls}}& \multirow{2}{2cm}{Signature Verification} & \multirow{2}{5.5cm}{Transport protocol needs to be modified before using to validate server information} &\multirow{2}{*}{$\times$} &\multirow{2}{*}{\CheckmarkBold} & \multirow{2}{*}{$\times$}&\multirow{2}{*}{$\times$} &\multirow{2}{*}{$\times$} \\
     &  & & & & & &\\ 
\ChangeRT{0.05pt}

   \multirow{2}{2cm}{\cite{adler2018astraea}}& \multirow{2}{2cm}{Incentivize Voting} & \multirow{2}{6cm}{Nash equilibrium can be satisfied in rational participants but does not guarantee authenticity} &\multirow{2}{*}{$\times$} &\multirow{2}{*}{$\times$} & \multirow{2}{*}{\CheckmarkBold} &\multirow{2}{*}{$\times$} &\multirow{2}{*}{$\times$} \\
   &  & & & & & &\\

\ChangeRT{0.05pt}

  \multirow{2}{2cm}{\cite{meter2022}} & \multirow{2}{2cm}{Sidechain Extension} & \multirow{2}{5.5cm}{It increases the scalability and throughput as a chain, but it can also be attacked} &\multirow{2}{*}{$\times$} &\multirow{2}{*}{$\times$} &\multirow{2}{*}{$\times$} &\multirow{2}{*}{$\times$} & \multirow{2}{*}{\CheckmarkBold} \\
   &  & & & & & &\\
\ChangeRT{0.05pt}
  
  \multirow{2}{1.5cm}{\cite{makerdao2020}}& \multirow{2}{2cm}{Committee Voting }& \multirow{2}{6cm}{Ability to handle lots of request transactions, but they cannot be processed immediately} &\multirow{2}{*}{$\times$} &\multirow{2}{*}{$\times$} &\multirow{2}{*}{$\times$} &\multirow{2}{*}{\CheckmarkBold} &\multirow{2}{*}{$\times$} \\
   &  & & & & & &\\

\ChangeRT{0.05pt}

  \multirow{2}{1.5cm}{\cite{zhang2020deco}}& \multirow{2}{2.5cm}{Zero-Knowledge Proof}& \multirow{2}{6cm}{The price request can be proved to be correct and does not need to modify the server}&\multirow{2}{*}{$\times$} &\multirow{2}{*}{$\times$} &\multirow{2}{*}{$\times$} & \multirow{2}{*}{\CheckmarkBold}&\multirow{2}{*}{$\times$} \\
  & & & & & & &\\
  \ChangeRT{0.05pt}

  \multirow{2}{2cm}{\cite{adams2021uniswap}} & \multirow{2}{2cm}{Symbolic Equation} & \multirow{2}{5.5cm}{Trading frequently by AMM, but rivals can profit from economic attacks}&\multirow{2}{*}{\CheckmarkBold} &\multirow{2}{*}{$\times$} &\multirow{2}{*}{$\times$} &\multirow{2}{*}{$\times$} &\multirow{2}{*}{$\times$}\\
   &  & & & & & &\\ 
  \ChangeRT{0.05pt}

  \multirow{2}{1.5cm}{\cite{wang2021promutator}}  & \multirow{2}{2cm}{Track Comparison} &\multirow{2}{5.5cm}{The modified EVM runs historical data to detect if the oracle is vulnerable to attacks}&\multirow{2}{*}{$\times$} &\multirow{2}{*}{$\times$} &\multirow{2}{*}{$\times$} &\multirow{2}{*}{$\times$} & \multirow{2}{*}{\CheckmarkBold} \\
  &  & & & & & &\\
  \ChangeRT{0.05pt}

  \multirow{2}{1.5cm}{\cite{provable2022}}& \multirow{2}{2cm}{TEE} & \multirow{2}{6cm}{Several trusted computations used to prove the data and transport it to the smart contract} &\multirow{2}{*}{$\times$} &\multirow{2}{*}{\CheckmarkBold} &\multirow{2}{*}{$\times$} &\multirow{2}{*}{$\times$} &\multirow{2}{*}{$\times$} \\
   &  & & & & & &\\ 

\ChangeRT{0.7pt}
  \end{tabular}
  \vspace{-1ex}
  \label{tab:oracle_solutions}
\end{table*}
Due to the necessity for off-chain asset information such as pricing, there is an expanding demand for superior oracles~\citep{RN68}. There have been decentralized and centralized related studies for this event, as summarized in Table \ref{tab:oracle_solutions}, where the 'C' means centralized, 'D' means decentralized, 'On' represents the method is on-chain, and 'Off' is off-chain.

Town Crier (TC)~\citep{zhang2016town} employs SGX technology, a Trusted Execution Environment (TEE) offered by Intel, to address the issue of safe communication between blockchain smart contract $SC=(SC_1,SC_2)$ and the network layer. We assume that TC has one extra TCP layer than the HTTPS network protocol for providing dependable data sources. When $SC_1$ launches a transaction to $SC_2$, the program $prog$ created in preparation in TC receives the transaction's datagram request, then obtains the data source through the external network's HTTPS protocol, and finally delivers the request with a digital signature to the requester $SC_1$. It isolates the malicious network operation from the process running on the host.

Uniswap~\citep{adams2021uniswap} automatically reconciles the required cryptocurrency pricing information by using the Automated Market Maker (AMM), using an on-chain smart contract to set up the symbolic equation. We assume a centralized asset container $pool$ with a large number of two cryptocurrencies $a$ and $b$, which correspond to stocks of $(a)$ and $(b)$. Uniswap allows $(a)\times(b)=k $, where $k$ is a constant. The decrease in value of $a$ caused by obtaining a loan contributes to the expansion of $b$.

TLS-N~\citep{ritzdorf2017tls} is an extension of Transport Layer Security (TLS). TLS generates a network encrypted channel for data exchange to ensure that the adversary does not access the data of the conversation between the two parties. TLS-N is proposed to solve the requirement of the third trusted party. It starts evidence generation and collection after the traditional handshake, which means that the evidence generator signs the handshake state with the private key immediately after the handshake. And TLS-N records all the handshake operations. All handshake records and proof signatures are utilized to assure non-repudiation of the conversation.

Provable~\citep{provable2022} is similar to Chainlink~\citep{chainlink2022} in that it sends currency information to Ether via operational nodes. The distinction is that the former uses trusted computing to ensure the accuracy of information across the network and transport layers, while the latter operates outside the chain. The latter uses median value computations to store multi-party data in the decentralized oracle network, ensuring safe data exchange. In addition, it is essential to note that Chainlink uses a reputation system to encourage each node.

Astraea~\citep{adler2018astraea} provides a voting oracle approach using the game analysis. It selects the submitters of the funding allocation scheme, the voters who vote on the funding, and the validators who verify the correctness of the scheme. The benefit of employing this strategy is that it guarantees the accuracy of the data for reasonable participants, despite not ensuring their legitimacy.

Meter~\citep{meter2022} is a sidechain built in parallel to the blockchain using PoW consensus, increasing the main chain's throughput and scalability. The cryptocurrency MTR on Meter is an autonomous and distributed coin passed to other blockchains, such as Ethereum, through sidechain technology, making the price of the cryptocurrency stable. It improves not only the performance but also expansion.

MakerDAO~\citep{makerdao2020} takes various asset prices from multiple oracle sources, then the committee votes to select one set of trusted feeds and pass them to the smart contract. However, to prevent the attacked information from being uploaded to the chain, the oracle security mechanism delays the input of the prices for one hour after being obtained. The committee ultimately decided to establish an emergency oracle in order to defrost a malicious oracle and prevent an oracle crash.

DECO~\citep{zhang2020deco} is similar to TLS-N, they are both interested in evidence generation. TLS-N needs advanced modifications to the server protocol to generate a proof. In contrast, DECO generates a proof in this Section using the zero-knowledge proof, which means there is no a revelation of encryption keys. $P$ and $V$ are given a shared key by the prover $P$, the verifier $V$, and the TLS server $S$. $P$ initiates a query request using the shared key, $S$ responds and transmits data to $P$. $V$ is responsible for recognizing the request, responding to it, and then announcing its shared key. $P$ verifies the returned data in conclusion.

ProMutator~\citep{wang2021promutator} detects whether an oracle is vulnerable by analyzing normal and abnormal transactions against the price oracle. The transactions of price oracle attacks are executed in the modified EVM according to predefined rules. And then, the comparison of the original and mutating traces generates reports, which analyze the differences.

\subsubsection{Wallet Optimization Schemes}
Users initiate a transaction and sign it using the key pair, the assets in the account are risky when the key leaks to an adversary. In Table \ref{tab:wallet_optimization}, some studies~\citep{RN64,RN80,RN74,RN77,dai2018sblwt,rezaeighaleh2019deterministic,jian2021securing,han2021efficient} proposed specific solutions for wallet management and wallet architecture.

\begin{table}[h]
\setlength{\abovecaptionskip}{0cm} 
\setlength{\belowcaptionskip}{-0.2cm}
    \centering
    \caption{Summarization of Wallet Security Optimization Schemes. }
    \renewcommand{\arraystretch}{1.3} 
\footnotesize
    \begin{tabular}{|p{2cm}| p{5cm}| }
        \ChangeRT{0.7pt}
        \makecell[l]{\textbf{Reference}} & \makecell[l]{\textbf{Key Scheme}} \\
        
        \ChangeRT{0.05pt}
        \multirow{2}{2.4cm}{\cite{RN64}} & \multirow{2}{5cm}{Improving collaborative key generation and signature }\\
        &\\
        \ChangeRT{0.05pt}
        \multirow{2}{2cm}{\cite{RN80}} & \multirow{2}{5cm}{Combination with a cold and hot wallet for privacy} \\
        &\\
        \ChangeRT{0.05pt}
        \multirow{2}{1.5cm}{\cite{RN74} }& \multirow{2}{5cm}{A practical way to public key encryption} \\
        &\\
        \ChangeRT{0.05pt}
        \multirow{2}{2cm}{\cite{RN77}} & \multirow{2}{5cm}{A p2p wallet scheme with a routing protocol} \\
        &\\
        \ChangeRT{0.05pt}
        \multirow{2}{2cm}{\cite{dai2018sblwt}} & \multirow{2}{5cm}{A lightweight wallet based on TEE} \\
        &\\
        \ChangeRT{0.05pt}
        \multirow{2}{2cm}{\cite{rezaeighaleh2019deterministic}} & \multirow{2}{5cm}{A new scheme for creating sub-wallet keys} \\
        &\\
        \ChangeRT{0.05pt}
        \multirow{2}{2cm}{\cite{jian2021securing}} & \multirow{2}{5cm}{The shared key generated by T-ECDSA for signing} \\
        &\\
        \ChangeRT{0.05pt}
        \multirow{2}{2cm}{\cite{han2021efficient}} & \multirow{2}{5cm}{Combine signatures into one based on the bloom filter} \\
        &\\
        \ChangeRT{0.7pt}
    \end{tabular}
    \vspace{-2ex}
    \label{tab:wallet_optimization}
\end{table}

According to~\citep{RN64}, existing hardware wallets migrated from the PC wallet architecture, resulting in a bad design that does not fundamentally fix the problem when just utilizing authentication and communication encryption. Interactive authentication adds several signatures and keys to the original wallet structure, which prevents attackers from manipulating the keys for transactions using a malfunctioning wallet.

Two applications for Android that in~\citep{RN80} were developed to provide privacy protection when used in conjunction with software and hardware. It has a cold wallet that stores the private keys in the form of QR codes and a hot wallet that can use to send transactions.

  
\cite{RN74} presented a new practical way of public key cryptography deployment. It formalizes the user's interactions with the management server $m$, the central server $c$, and the proxy $p$. This method provides five protocols, the first of which initializes all parameters in preparation for secure channels and verification operations, such as producing keys and verifying digital signatures. The second process is registration, in which $u$ is bound to the $m$ and produces valid login credentials. The registration data is then sent to $c$. The third is a backup, and the fourth is a verification that is utilized to perform a transaction in $p$. The final service provided by it is wallet recovery with the help of registration and backup.

Utilizing a routing protocol, \cite{RN77} converted the C/S architecture into a Peer-to-Peer (P2P) structured wallet management scheme. It then distributes the data using a Peer-to-Peer network after solving a problem involving multiple constraints simultaneously. In addition, it proposes a new strategy for key sharing called SKN, which aims to increase the availability of keys. In order to ensure the decentralization, where the same key is not stored on the same node. In addition, a streaming network is an example of a Fully Connected Network (FCN) through which users can communicate.


\cite{dai2018sblwt} came up with the SBLWT, a lightweight wallet architecture based on TEE, and designed it. The insecure storage module is present in the trusted environment. And it is present in the trusted execution environment (TEE), where users can just get the read access to the encrypted block header. On the other hand, encrypted messages can be proactively stored in insecure spaces, ensuring security between message exchanges.

Unlike traditional forwarding wallets, \cite{rezaeighaleh2019deterministic} shared the keys of the main wallet and sub-wallets to enable the generation of various sub-wallet addresses in a transaction. And the address generation process in the sub-wallets is designed to secure its keys. Most importantly, there is no need to back up the sub-wallets, the master-sub structure can derive the sub-wallets from the master wallet.

To defend against a single point of failure, \cite{jian2021securing} proposed a scheme with Threshold-based ECDSA (T-ECDSA). When the number of participants is within the threshold, they create shared private keys to sign transactions. Participants outside the threshold take turns signing transactions. There is also another program~\citep{han2021efficient} that uses T-ECDSA to design wallets. It combines multiple signatures into one signature. However, the design of the bloom filter protects the information of the participants on a small scale.

\subsection{P2P Network Optimizations}
The transactions initiated by each node in Ethereum are transmitted through P2P networks to achieve self-governance without a third party; however, the lack of authentication and other features leads to a series of attacks, such as the eclipse attack~\citep{RN103,RN102,RN105,henningsen2019eclipsing} in Table \ref{tab:p2p_optimization}. An information eclipse attack is one in which the attacker removes nodes from a network to restrict access to the information they contain.

\begin{table}[htbp]
\setlength{\abovecaptionskip}{0cm} 
\setlength{\belowcaptionskip}{-0.2cm}
    \centering
    \caption{Summarization of Network Security Optimization Schemes.}
    
    \renewcommand{\arraystretch}{1.3} 
\footnotesize
    \begin{tabular}{|p{2.5cm}| p{4.5cm}| }

        \ChangeRT{0.7pt}
        \makecell[l]{\textbf{Reference}} & \makecell[l]{\textbf{Key Scheme}} \\
        \ChangeRT{0.05pt}
         \multirow{2}{3cm}{\cite{RN103}} &\multirow{2}{4.5cm}{Sending requests to multiple peers} \\
         & \\
        \ChangeRT{0.05pt}
       \multirow{2}{2.5cm}{\cite{RN102}} & \multirow{2}{4.5cm}{Ensure that node ids always exist in the query table} \\
       & \\
        \ChangeRT{0.05pt}
        \multirow{2}{1.5cm}{\cite{RN105}} & \multirow{2}{4.5cm}{Analyze packets using a random forest model} \\
        & \\
        \ChangeRT{0.05pt}
       \multirow{2}{3cm}{ \cite{henningsen2019eclipsing}} & \multirow{2}{4.5cm}{Increasing the default number of peers} \\
        &\\
        \ChangeRT{0.7pt}

    \end{tabular}
    \vspace{-3ex}
    \label{tab:p2p_optimization}
\end{table}

\cite{RN103} proposed a novel eclipse bug, when the block height is $n$, a malicious node can obtain the ${(n+1)}_{th}$ block by preventing a regularly functioning node $N$ from receiving it. It invalidates subsequent blocks, even if node $N$ may receive them. However, it offers some countermeasures. If there is a block request that is not corresponding, the block is requested from multiple peers instead of just one peer. Through multi-party collaborative governance, this approach can solve denial of service attacks caused by eclipse attacks.

However, \cite{RN102} suggested a series of protection methods against eclipse attacks on Ethereum, two of which are also adopted by Geth. When a node restarts, the client's \texttt{seeding} is triggered every hour, or \texttt{lookup()} is called on an empty \texttt{table} which stores the information in \texttt{memory}, but the \texttt{seeding} is available only if the \texttt{table} is empty. However, node IDs should always be inserted into the \texttt{table} to prevent attacks. Specifically, Geth runs a \texttt{lookup()} on three random targets during \texttt{seeding} to add more legitimate nodes from the \texttt{db}, which stores the information on disk to the \texttt{table} to prevent attackers from inserting their node IDs into an empty \texttt{table} during \texttt{seeding}.

\cite{RN105} provided the ETH-EDS model for analyzing packets and identified three features in packets for the Random Forest (RF) model, which is a classification model based on decision trees. It identifies malicious nodes that are isolating users, allowing users to defend their networks in real time.

\cite{henningsen2019eclipsing} proposed an optimization scheme for the false friend eclipse attack that exists in Geth. The attack can be deployed with just 2 IPs and loaded on the node immediately. The peer is the number of network nodes that can be connected, default by 25. It is similar to \cite{RN103} in that it increases the peers to 50 to increase the probability of survival of spare nodes.

\subsection{Consensus Layer Optimizations}
The consensus layer and the incentive layer are interdependent. The design of the consensus mechanism directly affects the behavior of miners. Although many consensus mechanisms have been proposed, there is little regulation of the consensus or incentive levels. Table \ref{tab:consensus_optimization} summarizes the following optimized solutions.

\cite{yang2021finding} used the tool called Fluffy to discover two vulnerabilities, "Shallow copy" and "Ether Shift" in Table \ref{tab:consensus_bugs}. It first picks test cases and then changes the transactions in those cases. Next, it puts these transactions into multiple EVMs and collects all the state and coverage at the end of the execution. Finally, it repeats the above steps until the state no longer changes.

\begin{table}[ht]
\setlength{\abovecaptionskip}{0cm} 
\setlength{\belowcaptionskip}{-0.2cm}
    \centering
    \caption{Summarization of Consensus Security Optimization Schemes.}
    \renewcommand{\arraystretch}{1.3} 
\footnotesize
    \begin{tabular}{|p{2cm}| p{5cm}|}
        \ChangeRT{0.7pt}
        \makecell[l]{\textbf{Reference}} & \makecell[l]{\textbf{Key Scheme}} \\
        
        \ChangeRT{0.05pt}
        \multirow{2}{2cm}{\cite{yang2021finding}} & \multirow{2}{5cm}{Execute variant transactions with several EVMs} \\
        & \\
        \ChangeRT{0.05pt}
        \multirow{2}{2cm}{\cite{RN1}} & \multirow{2}{5cm}{Build the graph and analyze it to find fork attacks} \\
        & \\
        \ChangeRT{0.05pt}
        \multirow{2}{2cm}{\cite{Evmlab2017}} & \multirow{2}{5cm}{Randomly select an EVM version to test transactions} \\
        & \\
        \ChangeRT{0.05pt}
        \multirow{2}{2cm}{\cite{libFuzzer2017}} & \multirow{2}{5cm}{Fuzz engine for parallel extraction of functions} \\
         & \\
        \ChangeRT{0.05pt}
       \multirow{2}{1.5cm}{\cite{fu2019evmfuzzer}} &\multirow{2}{5cm}{Test EVMs with contracts and analyze results}\\
        & \\
        \ChangeRT{0.7pt}
    \end{tabular}
    \vspace{-2ex}
    \label{tab:consensus_optimization}
\end{table}

\begin{figure}[htp]
\setlength{\abovecaptionskip}{0cm} 
    \centering
    \includegraphics[width=8cm]{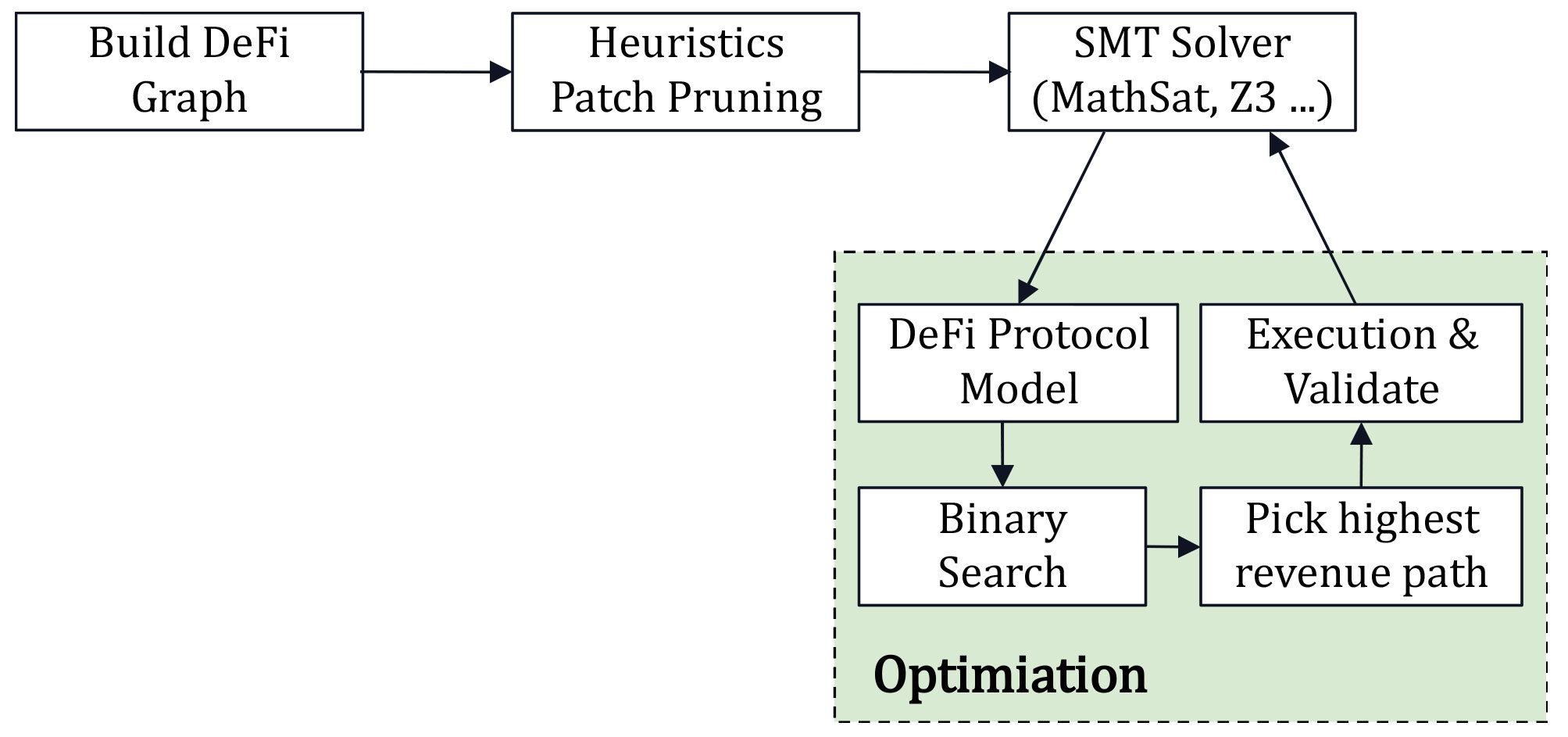}
    \caption{Diagram of \textsc{DefiPoser} Core Process.}
    \vspace{-4ex}
    \label{fig:DefiPoser}
\end{figure}

As described in Section \ref{sec:con_mec_v}, fork attacks might affect the security of blockchain in terms of consensus mechanism, \cite{RN1} developed \textsc{DefiPoser} to monitor fork behaviors. Figure~\ref{fig:DefiPoser} shows the process of \textsc{DefiPoser}. It prunes the patches after building the DeFi graph based on the heuristic approach~\citep{dwivedi2020security} and then does a greedy search of the negative cycle in the directed transaction flow graph, which means finding all possible profitable cycles in the trade flow graph, to detect arbitrage transactions in cyclic or more complicated scenarios. A binary search of all the paths finds the most profitable one. If it is within the quantization threshold quantified by the Markov decision process, there is a chance to motivate a fork attack by miners using MEV.

Ethereum exists an EVMLab~\citep{Evmlab2017} library for interacting with Ethereum virtual machines (EVMs), which are officially used to analyze the bytecode of smart contracts. In this way, one version of the EVM is randomly selected for a single transaction on a smart contract, and then finds bugs.

Another library called LibFuzzer~\citep{libFuzzer2017}, was developed based on a toolchain LLVM written in C++. It generates $N$ concurrent processes for functions in the contract and randomly assigns subsets to them, where one of the subsets will merge its generated corpus into the main set in the end. These corpora are used to find bugs using fuzz.

EVMFuzzer~\citep{fu2019evmfuzzer} is a tool for testing and evaluating EVM. It takes the target EVM and its API as input and then creates an execution environment for them to test and evaluate the EVM. In this process, multiple EVMs receive some of the same quality contracts that have been selected and output the results in the same format. It discovered a DoS attack on Geth, which was recorded as CVE-2018-19184.
 
\subsection{Smart Contract Security Optimizations}
The smart contract, a component of the DeFi project connecting the data layer and the application layer, can modify the state of a transaction. It may lead to errors and be utilized by attackers. Consequently, it is critical to strengthen the security of contracts.
\begin{table*}[width=18cm, pos=htbp]
\setlength{\abovecaptionskip}{0cm} 
\setlength{\belowcaptionskip}{-0.2cm}
\centering
\caption{Summarization of Methods for Smart Contract Vulnerabilities Detection.}
\footnotesize
\begin{tabular}{|p{2cm} |p{3cm}| p{4cm}| p{6cm}|} 
\ChangeRT{0.7pt}
      \textbf{Reference} & \textbf{Key Technologies} & \textbf{Target Vulnerability} & \textbf{Features}    \\


\ChangeRT{0.5pt}

  \multirow{4}{1.5cm}{\cite{RN26}}
        & \multirow{4}{3cm}{Symbolic Execution}  &   Transaction State Dependency 
        & \multirow{4}{6cm}{Control Flow Graph (CFG) Construction\\Symbolic Execution\\Core Analysis} \\
        & &Block Info Dependency &\\  
        & &Unhandled Exception &\\ 
        & &Reentrancy & \\
\ChangeRT{0.05pt}

   \multirow{2}{1.5cm}{\cite{RN81}}         
        &\multirow{3}{3cm}{Machine Learning\\Static Analysis}
        &  \multirow{3}{4cm}{Ponzi Scheme}
        & Obtain account features from transactions\\
    & & &Obtain code features from opcodes\\
    & & &XGBoost\\
  \ChangeRT{0.05pt}

  \multirow{1}{2cm}{\cite{amani2018towards}} 
        &\multirow{2}{3cm}{Formal Verification}
        & \multirow{2}{4cm}{Logical Vulnerabilities} 
        & Formal EVM extensions with Isabelle/HOL\\
        & & &Logical verification at bytecode level \\
        
\ChangeRT{0.05pt}

   \multirow{6}{1.5cm}{\cite{RN39}}
        & 
                &Greedy Contract
                &\multirow{6}{6cm}{Symbolic Execution\\GRU\\Fully Connected Network (FCN)\\Fuzz}\\
        & &Ether Leaking &  \\
        &Machine Learning &Suicidal Contract & \\
        &Fuzz &Block Info Dependency &\\
        & &Unhandled Exception &\\
        & &Controlled Delegatecall &\\
\ChangeRT{0.05pt}

  \multirow{5}{1.5cm}{\cite{RN85}}
        &\multirow{5}{3cm}{Machine Learning}
                & Overflow
                &\multirow{5}{6cm}{Clone Detection\\Code Embedding\\Similarity\\Checking}\\     
        & &Block Info Dependency &\\
        & &Reentrancy &\\
        & &Greedy Contract & \\
        & &Bad Randomness &\\
\ChangeRT{0.05pt}

  \multirow{3}{1.5cm}{\cite{xue2020cross}}
        & Static Analysis 
                & \multirow{3}{4cm}{Reentrancy} 
                & \multirow{3}{6cm}{Cross-contract CFG\\Static Taint + PPT}\\
        &Path Protection & & \\
        &Technology (PPT) & & \\
  \ChangeRT{0.05pt}

  \multirow{3}{1.5cm}{\cite{so2020verismart}}
        & \multirow{3}{3cm}{Static Analysis} 
        & Arithmetic Bug
                & \multirow{3}{6cm}{Insert assertions generate many queries and invariants, queries are validated with Solver}\\
            & &Assertion Violation & \\
            & &ERC-20 Standard Violation & \\
  \ChangeRT{0.05pt}

  \multirow{3}{1.5cm}{\cite{RN83}}
        &\multirow{3}{3cm}{Machine Learning \\Symbolic Execution}
        & \multirow{3}{4cm}{Ether Leaking \\Suicidal Contract}  
        & Hunt sequences with symbol execution\\
    & & &Train language model with sequences\\
    & & &Guide symbolic execution with the model\\
\ChangeRT{0.05pt}

  \multirow{4}{2cm}{\cite{huang2021hunting}}
        &\multirow{4}{3cm}{Machine Learning} 
            &Overflow 
            &CFG Construction \\
        & &Reentrancy &Slicing\\
        & &Unexpected Permission Check &Graph Embedding\\
        & &Bad Randomness &Similarity Checking \\
\ChangeRT{0.05pt}

  \multirow{8}{1.5cm}{\cite{RN28}}
        & \multirow{8}{3cm}{Static Analysis} 
        & Transaction State Dependency  
        & \multirow{8}{6cm}{CFG Construction\\Feature Detection\\Core analysis}\\
    & &DoS Under External Influence &\\
    & &Strict Balance Equality &\\
    & &Reentrancy &\\
    & &Nested Call &\\
    & &Greedy Contract &\\
    & &Unchecked External Calls &\\
    & &Block Info Dependency &\\
\ChangeRT{0.05pt}

  \multirow{10}{1.5cm}{\cite{RN41} }   
        &\multirow{10}{3cm}{Static Analysis\\Fuzz}
        &Assertion Violation
            &\multirow{10}{6cm}{Collection of bytecode data streams\\Select semantic information as seeds\\Fuzz based on data stream feedback}\\
    & &Unexpected Permission Check & \\
    & &Block Info Dependency & \\
    & &Ether Leaking & \\
    & &Greedy Contract & \\
    & &Arithmetic Bug & \\
    & &Unchecked External Call & \\
    & &DoS Under External Influence & \\
    & &Reentrancy & \\
    & &Suicidal Contract & \\
\ChangeRT{0.05pt}

  \multirow{3}{2cm}{\cite{nam2022formal}}     
        & \multirow{3}{3cm}{Formal Verification\\Game Theory}  
        &\multirow{3}{4cm}{Interactive Contract Vulnerabilities}
            &Convert source files to MCMAS files\\
    & & &Construct contract game structure\\
    & & &Examine ATL properties in the structure\\
  \ChangeRT{0.05pt}

  \multirow{5}{1.5cm}{\cite{jin2022exgen}}
        &\multirow{5}{3cm}{LLVM IR\\Symbolic Execution }  
         &Arithmetic Bug  
            &Convert the source code to LLVM IR\\
    &  &Reentrancy &Locating Vulnerabilities\\
    & &Suicidal Contract &Collect ordered sets of transactions\\
    & &Unchecked External Calls &Symbolic Execution\\
    & & &Verification\\

\ChangeRT{0.7pt}
  \end{tabular}
  \label{tab:smartcontract_solutions}
\vspace{-1ex}
\end{table*}

\subsubsection{Smart Contract Vulnerability Detection}

In this Section, we will discuss the methods presented in Table \ref{tab:smartcontract_solutions} that were suggested for detecting the vulnerabilities in the smart contract. Malicious actors can exploit these flaws. Formal verification, symbolic execution, and machine learning are some of the technological tools~\citep{RN28,RN26,amani2018towards,RN39,RN85,xue2020cross,so2020verismart,RN83,huang2021hunting,RN41,nam2022formal,jin2022exgen} that have been the subject of a significant amount of research conducted to disclose contract vulnerabilities.

\begin{figure}[htp]
    \centering
    \includegraphics[width=8cm]{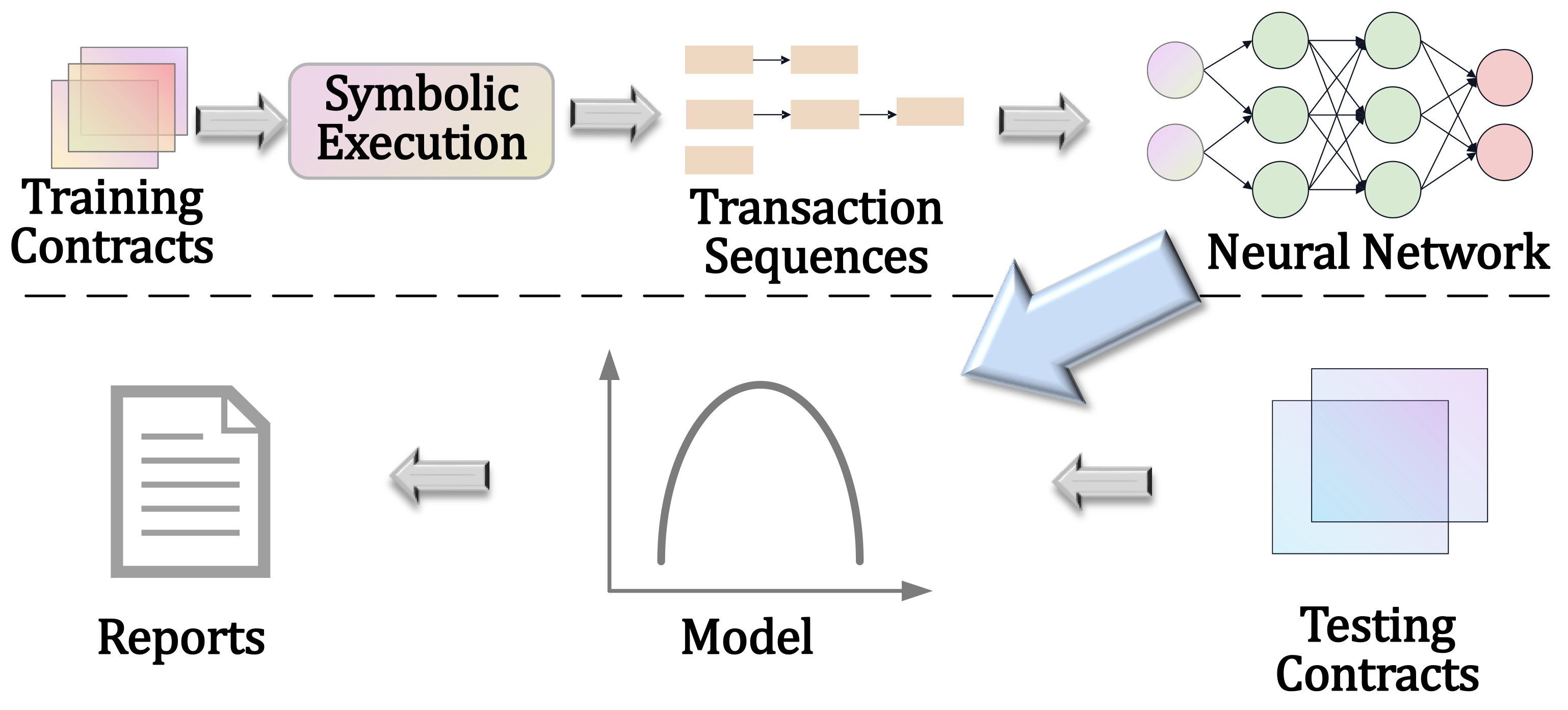}
    \caption{Schematic Diagram of \textsc{ILF} Process Framework.}
    \vspace{-3ex}
    \label{fig:ILF}
\end{figure}

Combined with dynamic testing extends the ability of symbolic execution techniques to detect unknown vulnerabilities, thus improving the robustness of programs. Figure~\ref{fig:ILF} shows an overview of \textsc{ILF}~\citep{RN39} that combines fuzzing, machine learning, and symbolic execution. The system used the symbolic execution for a portion of the contracts to generate transaction sequences as the training dataset for a new model consisting of GRU, which is a type of neural network and a fully connected network so that the model can learn the fuzzing in the state after the symbolic execution to test contracts with high coverage.

As our best known, Oyente~\citep{RN26} is the first detection tool using symbolic execution for smart contracts. It examines the logic of the contract code and generates control flow graphs. It then instructs the Ethereum initial state simulation run to construct feasible data flow operations. After that, the appropriate analysis methods find a variety of different vulnerabilities.

\cite{RN81} counted the frequency of opcodes in the contract and analyzed the features of malicious accounts through the ether flow graph. Then, for vulnerability identification, the XGBoost model~\citep{chen2016xgboost}, which is a gradient model based on the decision tree, was built using the features. This strategy can be utilized before contract deployment because it does not need attributions about the transaction.

\cite{amani2018towards} formalized the EVM in terms of bytecode using Isabelle/HOL and built a double verifier logic on reasoning about the program. They then demonstrated the safety of the system. However, owing to the development of Ethereum, it cannot describe the complete semantics of the smart contract in several versions.

DefectChecker~\citep{RN28} uses the commands provided by Geth to disassemble the contract bytecode into opcodes and then split the opcodes into multiple base blocks, executing different instructions for each block and defining eight types of rules to detect vulnerabilities. It analyzes specific features of the vulnerability, and then different generic specifications are set for detecting the vulnerability based on \cite{RN34}.

SmartEmbed~\citep{RN85} consists of two main aspects, structured code embedding that converts code into word vectors and similarity checking that can detect the similarity of different vectors. It marks Solidity codes, and a new code embedding is generated using a word embedding techniques, and finally, bugs can be found within a threshold by comparing the similarity between vectors.

Clairvoyance~\citep{xue2020cross} designed several path protection techniques for reentrancy vulnerability and used taint analysis techniques to reduce the phenomenon of false positives from other tools. More importantly, the lightweight approach allows analysis of cross-contract behavior.

Up to May 2022, the compiler of Solidity has over 90 versions, each with significant updates~\citep{RN43}. As a result, the compiled bytecodes with the same logic are diverse and noisy. To solve this problem, \cite{huang2021hunting} labelled the data and reordered the opcodes. This process ignores all irrelevant instructions, then analyzes the bytecode execution process and slices the data by the label to reduce the noise impact of meaningless code. Subsequently, this method uses an unsupervised graph embedding algorithm to deal with the smart contracts, each slice of code encoded as a vector and their similarity compared for vulnerability detection.

Smartian~\citep{RN41} just started statically analyzing the contract bytecode and collecting the data stream. The seed pool initialization predicts the sequence of transactions in some data stream and considers them as seeds for initialization. The seeds, the sequences of generated transactions, are then used to guide the fuzzy logic of the data flows.

Another methodology similar to ILF uses high-quality transaction sequences to direct symbolic execution in order to discover susceptible contracts. SmarTest~\citep{RN83} based on VeriSmart~\citep{so2020verismart}, first performs symbolic executions on the contracts in the training dataset, each lasting long enough to gather all the fragile transaction sequences, and then utilizes these fragile transaction sequences as the collection of training sequences for the language model. The goal of training the language model is to create a training corpus $Y$ from which counts of $n$ tuples are gathered to guide the symbolic execution to find fragile transaction sequences.

\cite{nam2022formal} examined smart contracts using the Alternating-time Temporal Logic (ATL) model, a formal verification approach that determines if smart contracts in the Ethereum meet certain features. Additionally, it converts Solidity to MCMAS, an ATL checker that requires input from the user. It enables developers to validate the attributes they would like to examine.

EXGEN~\citep{jin2022exgen} transforms the contract source code or bytecode into an Abstract Syntax Tree (AST), which is subsequently converted into an LLVM representation of the unified form. Afterward, all restrictions are eliminated utilizing symbolic execution techniques. The system then employs a solver to resolve the constraints and establish the order of vulnerabilities. After that, the sequence is reviewed for reliability before being added to the blockchain.

\subsubsection{Smart Contract Operation Regulation}
\begin{figure}[htbp]
\setlength{\abovecaptionskip}{0cm} 
    \centering
    \includegraphics[width=6.5cm]{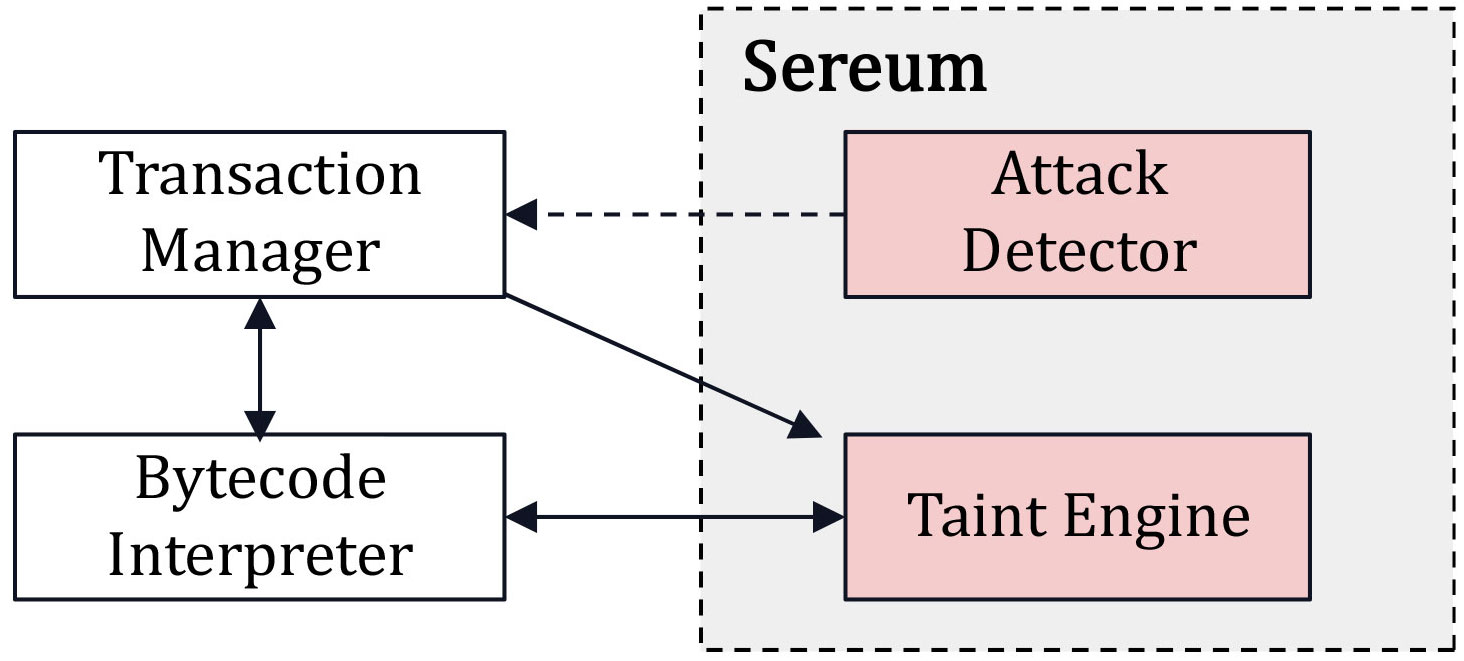}
    \caption{Diagram of \textsc{Sereum} System Architecture.}
    \vspace{-2ex}
    \label{fig:Sereum}
\end{figure}

Smart contracts can be more secure by detecting vulnerabilities, according to Torres et al.~\citep{RN87}. However, the number of assaults has not decreased, which indicates that contract regulation needs to be improved. It has been studied in~\citep{RN93,RN92,RN37,RN91,RN87,nguyen2021sguard} in Table \ref{tab:smartcontract_Regulations}, and we briefly introduce Sereum~\citep{RN93} in Figure~\ref{fig:Sereum}, a tool focused on runtime monitoring and verification of the reentrancy bug.

\begin{table*}[width=18cm, pos=ht]
\setlength{\abovecaptionskip}{0cm} 
\centering
\caption{Summarization of Methods for Smart Contract Vulnerability Regulation.}
\footnotesize
\renewcommand{\arraystretch}{1.3} 
\begin{tabular}{|p{2cm} |p{3.5cm}| p{3cm}| p{6.5cm}|} 
\ChangeRT{0.7pt}
      \textbf{Reference} & \textbf{Target Vulnerability} & \textbf{Key Technologies} & \textbf{Features}  \\


\ChangeRT{0.5pt}

  \multirow{3}{2cm}{\cite{RN93}} & \multirow{3}{3.5cm}{Reentrancy}
        & \multirow{3}{2cm}{Dynamic Taint Technique }
                        &\multirow{3}{6.5cm}{It traces the \texttt{storage} Variable and writes them into the lock during the transaction, when an attack is detected, the transaction would be rolled backed}
                \\
                &  & &\\
                &  & &\\
                
\ChangeRT{0.05pt}
  \multirow{4}{2cm}{\cite{RN92}} & 
        & \multirow{4}{3cm}{Domain-Specific Language (DSL)\\Vulnerability Patterns\\Voting Mechanism} 
                        & \multirow{4}{6.5cm}{\textsc{ÆGIS} modified the EVM to revert the codes written by DSL, which described the vulnerability patterns. Therefore, it can prohibit malicious control flow and data flow by comparing the patterns}
                \\
                &Reentrancy  &  & \\
                &\multirow{2}{3.5cm}{Unexpected Permission Check}  & & \\
                &   & & \\
\ChangeRT{0.05pt}
  
  \multirow{8}{2cm}{\cite{RN37}} & Reentrancy & \multirow{8}{3cm}{Online Framework}
                        &\multirow{8}{6.5cm}{\textsc{SODA} is an online framework that can be divided into two layers. The lower layer collects EVM information, and the higher layer provides interfaces for developing detection apps} \\
                                     &   Block Info Dependency& & 
                                 \\  & Strict Balance Equality & & 
                                 \\  &  Unchecked External Calls& & 
                                 \\  & \multirow{2}{3cm}{Unexpected Permission Check} & & 
                                 \\
                                 & & & 
                                 \\  
                                 & Missing Return & & 
                                 \\  & Greedy Contracts & & \\
  \ChangeRT{0.05pt}
  \multirow{5}{2cm}{\cite{RN91}} & 
            & \multirow{5}{3cm}{Patch}
                        &\multirow{5}{6.5cm}{It first makes the vulnerability detection utilizing other detection tools, and then to fix the contract, it rewrites bytecode. Finally, \textsc{EVMPatch} tests the fixed contract with the historical transactions to verify whether the patch is correct}
                \\  &Reentrancy    & & 
                \\  &Arithmetic Bug    & & 
                \\  &\multirow{2}{3.5cm}{Unexpected Permission Check}    & & 
                \\  &    & & \\
  \ChangeRT{0.05pt}
  \multirow{7}{2cm}{\cite{RN87}} & Reentrancy
        & \multirow{7}{3cm}{Mapping Knowledge Domain\\Dynamic Taint Analysis}
                        &\multirow{7}{6.5cm}{It extracts information from tracked transactions, and builds a graph with the nodes and transactions. Horus identified attacks with the graph and the queries, and finally loaded the tracing assets into a graph database}
               \\
                &  Arithmetic Bug & &\\
                &  Unchecked External Calls& &\\ 
                &  \multirow{2}{3cm}{Unexpected Permission Check} & &\\
                &&&\\
                &  \multirow{2}{3cm}{Unmatched ERC-20 Standard} & &\\
                &&&\\
  
  \ChangeRT{0.05pt}
  \multirow{4}{2cm}{\cite{nguyen2021sguard}} & Reentrancy 
                & \multirow{4}{3cm}{Automatic Contract Transformation}
                &\multirow{4}{6.5cm}{Several runtimes in \textsc{SGUARD} are used to identify control dependency, and it uses symbolic trace generations for each loop to get more data dependencies. When it finds and fixes bugs based on their definitions}
               \\
            &  Arithmetic Bug & & \\
            &  \multirow{2}{3cm}{Transaction State Dependency} & &\\
            &&&\\
\ChangeRT{0.7pt}
  \end{tabular}
  \vspace{-1ex}
  \label{tab:smartcontract_Regulations}
\end{table*}

The transaction manager converts all control flows into conditional jump instructions in the bytecode interpreter. The taint engine identifies data flows in conditional jump instructions, tagging storage variables as the key variables and writing into the lock. The attack detector detects the variables. If the modification occurs, the whole transaction rolls back to the point where the variable was marked, which is the starting point of the entire transaction.

Sereum defends online smart contracts from reentrancy attacks, but such systems are difficult to expand to additional weaknesses. EVMpatch~\citep{RN91} intended to address this issue by providing a bytecode rewriting engine that updates contracts currently on the chain, and the patch program is readily scaleable to other flaws. Automated analysis tools and vulnerability revelations detect and generate reports on vulnerabilities, and bytecode rewriters receive vulnerability reports and patch the contract at the byte level. The testing module then verifies that the patch will work with the previous transactions. After the test passes, the deployer uploads the patched contract to Ethereum.

Many tools only detect bugs but do not quantify and track stolen assets. To monitor stolen assets, the Eye of Horus~\citep{RN87} employs knowledge graph technology. It extracts data streams of vulnerable transactions using taint analysis. Then it analyzes the input data relationships and finds attacks from the generated logs. Finally, it obtains the attacker's addresses and timestamps, then loads the transactions into a graphical database for access to the asset flow.

ÆGIS~\citep{RN92} constructed some control flow and data flow patterns describing the vulnerabilities and modified the EVM so that it could revert to transactions against patterns during contract operation. Anyone can submit a pattern when discovering a new vulnerability. All voters in the chain then determine whether the pattern can be added to the list.

SODA~\citep{RN37} is an online detection framework for a smart contract that includes a manager, information collector, and logger. It provides registered and unregistered APIs to APPs and requires the APP to send operational information, block numbers, and different functions to the manager. The information collector collects all blocks, transactions, and contract structure information for vulnerability detection, and the logger issues an alert if it finds anomalies. This framework with detection APPs is compatible with multiple blockchains.

\begin{table*}[width=18cm, pos=ht]
\setlength{\abovecaptionskip}{0cm} 
\setlength{\belowcaptionskip}{-0.2cm}
\centering
\caption{Summarization of Methods for DeFi Optimization.}
\footnotesize
\renewcommand{\arraystretch}{1.3} 
\begin{tabular}{|p{2.5cm}| p{4cm} |p{3cm} |p{6cm}|} 
\ChangeRT{0.7pt}
      \textbf{Reference}  & \textbf{Target} & \textbf{Key Technologies} & \textbf{Features}   \\


\ChangeRT{0.5pt}

  \multirow{2}{2cm}{\cite{RN90}} & Data Flow Dependency
        & Symbolic Reasoning
                        &\multirow{1.5}{6cm}{Two sections track data flow dependencies and monitor malicious transactions respectively}\\
    &  Violation on Invariant  & Transaction Monitor& \\
  \ChangeRT{0.05pt}
  \multirow{3}{2cm}{\cite{RN55}}
  &\multirow{3}{4cm}{Price Manipulation Attack}
        &  Semantic Lifting
                        &\multirow{3}{6cm}{By collecting historical transaction data, a Cash Flow Tree (CFT) is constructed, which is used to recover high-level semantic information}
                \\
        &   &Pattern Detection &\\
        &  & &\\
  \ChangeRT{0.05pt}
    \multirow{3}{2cm}{\cite{xu2019anatomy}}
  &\multirow{3}{4cm}{Pump-and-Dump Attack}
        &  Machine Learning 
                        &\multirow{3}{6cm}{Random Forest and Generalized Linear Model (GLM) are employed for feature selection to prevent overfitting and regression classification to explain the process.}
                \\
        &   &Feature Engineering &\\
        &  & &\\
  \ChangeRT{0.05pt}
    \multirow{3}{2cm}{\cite{kamps2018moon}}
  &\multirow{3}{4cm}{Pump-and-Dump Attack}
        & Computed Anomaly Threshold
                        &\multirow{3}{6cm}{It specifies thresholds for detecting subtle abnormal changes and calculates temporal trends over time using windows. }
                \\
        &  & &\\
        &  & &\\
  \ChangeRT{0.05pt}
  \multirow{3}{1.5cm}{\cite{RN108}} & \multirow{3}{3cm}{Unexpected Code in Smart Contract}
         & Pricing
                        &\multirow{3}{6cm}{Tokens are used to relate membership relationships, and members work together to maintain contracts}
               \\
        & & \multirow{2}{3cm}{Decentralized Underwriting} &\\
        & & &\\
  \ChangeRT{0.05pt}
  \multirow{3}{2.5cm}{\cite{etherisc}}   & Crypto wallets
        &\multirow{3}{3cm}{Market-based Approach}
                        &\multirow{3}{6cm}{It is a centralized platform used for covering multiple risks, and if the user suffers enough damage, all losses can be refunded}
                \\
        &   Collateral & &\\
        &  & &\\
  \ChangeRT{0.05pt}
  \multirow{4}{2cm}{\cite{bright2021union}} & Protocol Failure
        & \multirow{4}{3cm}{Multiple-chain Covers Portfolio}
                        &\multirow{4}{6cm}{Integrate Various insurance and liquidity provision can improve the performance and maximize insurance revenues}
                \\
        &  Stablecoin& &\\
        &  Yield& &\\
        &  DEX & &\\
  \ChangeRT{0.05pt}
  \multirow{3}{2.5cm}{\cite{RN109}} & Derivatives
        & \multirow{2}{3cm}{Automated Squeeth Strategy}
                        &\multirow{3}{6cm}{Code audits and establishing reliable protocols for bounty payments}
                \\
        &  Options &  &\\
        &  & Liquidity Provider &\\
  \ChangeRT{0.05pt}

  \multirow{3}{1.5cm}{\cite{bridge2020mutual}}& Smart Contract
        & \multirow{3}{3cm}{Managed by DAO\\ Discretionary Coverage}
                        &\multirow{3}{6cm}{Coverage provision is added by users, so the coverage can be transparent, and there are many asset pools to share risks in the smart contract, stablecoin, and DEX}
               \\
        &  Stablecoin & &\\
        &  Decentralized Exchange (DEX) & & \\

\ChangeRT{0.7pt}
  \end{tabular}
  \vspace{-2ex}
  \label{tab:DeFi_optimation}
\end{table*}

\cite{nguyen2021sguard} proposed \uppercase{sGuard}, which evaluates control dependencies to discover malicious opcodes once all traces are enumerated at the bytecode level and located to external callers. Finally, the flaw is corrected through patching at the source code level. Correspondingly, it analyzes all types of data dependencies, including memory, storage, and stack. However, the limitation of this work is that there is no explicit number of iterations to obtain all data dependencies in smart contracts.

\subsection{Application Layer Optimizations}
There are some market manipulations at the application level that can lead to damage to user assets, but some research and works in Table \ref{tab:DeFi_optimation} exist to safeguard various applications, including optimization tools and insurance.

\subsubsection{Optimization Tools}
Although there is a correlation between the various layers, methods for lower layers can not fully recognize the attacks against the application layer. Some research\citep{RN90,RN55} makes contributions.

\begin{figure}[htp]
\setlength{\abovecaptionskip}{0cm} 
\setlength{\belowcaptionskip}{-0.2cm}
    \centering
    \includegraphics[width=8cm]{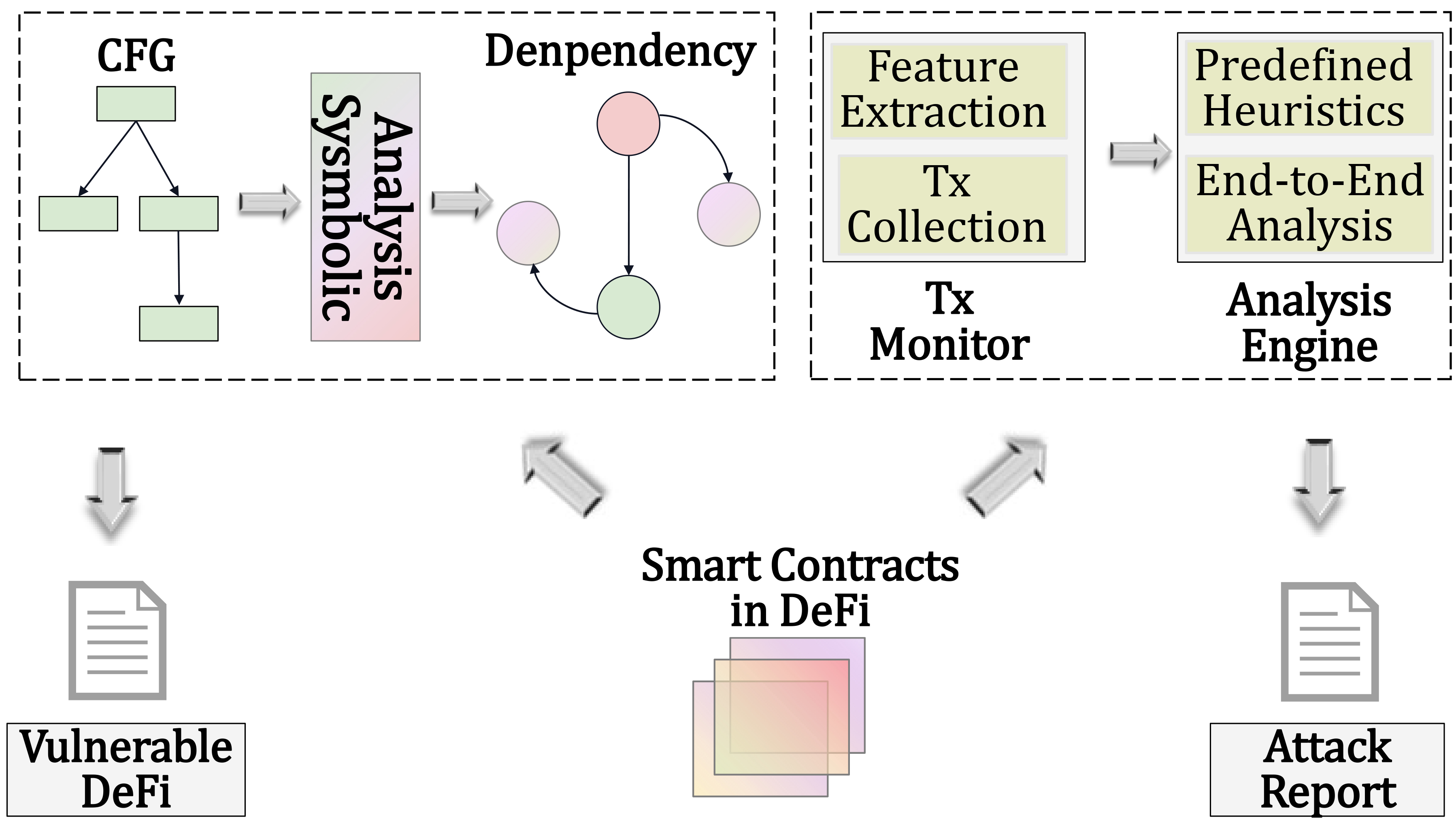}
    \caption{Diagram of \textsc{Blockeye} Core Process.}
    \vspace{-2ex}
    \label{fig:Blockeye}
\end{figure}

\cite{RN90} designed Blockeye to divide the detection work into two phases. In Figure~\ref{fig:Blockeye}, the first phase uses symbolic execution analysis in oracle to check whether state data streams are externally manipulated to detect vulnerable DeFi. During the second phase, transaction monitors under the chain collect transactions to extract the features and further analysis to monitor the attack.

DeFiRanger~\citep{RN55} is a price manipulation checker that first collects transactions on Ethereum and constructs a tree structure on the flow of cash. Then it defines the style based on DeFi attack behavior and recovers the low-level semantics to high-level semantics. Finally, the system detects attacks and analyzes them to generate a report based on the style and high-level semantics.

For the classification of Pump-and-Dump attacks, \citep{xu2019anatomy} employs machine learning techniques. A random forest model filters the features to prevent model overfitting and optimize the detection effect on anonymous data. A highly interpretable Generalized Linear Model (GLM) is employed to expedite the training process for large datasets.

\cite{kamps2018moon} discovers that most Pump-and-Dump transactions are local anomalies related to recent history. Given a period as a window and its average, it observes its trend by moving the window and finds local anomalies compared to recent history to detect such anomalies. However, this method may misdiagnose normal trades with large up-and-down scales.

\subsubsection{Insurance}
As the DeFi market expands, insurance is critical to ensuring its stability~\citep{RN108}. Our research divides risks in DeFi into market risks, technical risks, and credit risks. However, the enormous damages experienced by regular users result from technical or credit risks. So there require insurance systems to safeguard the properties of users, and they can be classified as centralized and decentralized.

For example, Smart Contract Cover, which provides insurance to smart contracts, is evaluated by the Nexus Mutual internal assessors to determine the cost of the insurance \citep{RN108}. It is self-governed by the members who own the NXM token of the pool built with "mutual", making the risk evenly shared.

Etherisc~\citep{etherisc} is a centralized platform that provides various insurance programs, such as wallets and collateral. As for cryptocurrency wallets, it protects them from hacking and theft. Another one refers to the protection of the price of the borrower's collateral within a certain range. For example, if the price of the collateral is reduced by more than 90\% due to a significant drop in the market, the insurance will pay lower price to the borrower.

Opyn~\citep{RN109}, which focuses on insurance for option trading products, enables users to choose options to hedge risks based on ERC20 tokens, and the protocol is automatically performed by smart contracts for multi-party governance. It also provides audit services and publishes all contracts, offering bounties for contract optimization.

Bridge Mutual~\citep{bridge2020mutual} allows users to add insurance to products of their choice, and the decentralization makes the entire coverage process transparent for insurance purposes. To achieve risk sharing, thousands of pools encompassing a variety of platforms, stable currencies, and exchanges are employed.

Bright Union~\citep{bright2021union} integrates most insurance in the industry and licenses to encrypt them to enable mutual coverage without a license. It can improve cost performance for higher complete coverage by combining various insurance. 


\section{Challenges and Future Directions}
\label{Challenges_Future}

\subsection{Challenges}
DeFi is a mechanism built through blockchain technology as the underlying technology that does not rely on the operation of a centralized service. The most significant advantage of DeFi is that funds are automatically traded through technical protocols. Thus it eliminates the need for human intervention, which increases liquidity in the market and thus facilitates financial transactions. However, DeFi still has challenges, and the following issues can be solved to facilitate the ecosystem's development better.

\textbf{(i)} The triangle problem in Blockchain remains a challenge. In Table \ref{tab:blockchains}, our analysis reveals that blockchains lose security when scalability or decentralization are enhanced. Various consensus algorithms prioritize each of the three directions. Scalability is sacrificed for security and decentralization in the Bitcoin blockchain. The trend of Ethereum ensures security and scalability but can only achieve partial decentralization. As the third generation, EOS blockchain technology prioritizes scalability and security using a fixed number of super nodes at the expense of decentralization.

\textbf{(ii)} Data issues remain a challenge. Technologies such as oracle are already being used to solve the problem of data synchronization, and wallets protect the key to ensure the encryption process. The improved transmission protocols are designed to secure the data transmission process. However,  under the complete autonomy of blockchain, the data cannot be modified after it is transmitted on the chain in principle. An error in the data transmission process or the data source will lead to problems in the system. The technology in these aspects is not mature enough. It cannot fully guarantee the correctness of data without affecting the other performance of the system, such as scalability and throughput.

\textbf{(iii)} Key technologies of blockchain are not mature enough. Starting with Bitcoin, the consensus mechanism is one of the fundamental technologies of blockchain. Even though it has few vulnerabilities, it still has a series of actions against consensus rule flaws that may result in forking. More importantly, the loss it causes is significant since forking is irreversible.
The smart contract is the basis of various DAPPs, and the vulnerabilities that exist in it are also endless. Due to the rapid development of the smart contract, there is still a lack of practical tools for potential vulnerabilities in it.

\textbf{(iv)} The technology features are not fully utilized, but there are just migrating from the existing mechanisms. The various existing economic instruments lead to technical indicators' failure. Since the DeFi applications are not fully integrated with the natural world, they currently relies on various virtual indicators for the prices of financial products. However, when some data beyond the normal range enters the system, it can lead to errors. Furthermore, since there is no trusted third party to guarantee DeFi, it has to rely on various protocol stacks, which leads to fragmentation and uncertainty of assets.

\textbf{(v)} The DeFi account has numerous issues. Accounts are the entities that perform transactions, and in the real world, users use them to send transactions to the blockchain world. The new users tend to choose DeFi apps with high transaction volumes and user activity. Many Bot Accounts in a DeFi application could create a risk of fraud against real users. Bot accounts are program-controlled accounts that initiate transactions to the DeFi app regularly, ensuring that the DeFi app is active. DeFi lacks the means to detect malicious bot accounts and is still in the traditional small-scale detection stage.
 
\textbf{(vi)} Some DeFi users engage in irresponsible financial conduct. DeFi evolved from traditional forms of finance, and it contains the same wrong behaviors in conventional finance, such as market manipulation and arbitrage. Most DeFi security works currently focus on researching smart contract vulnerabilities. It lacks research on the vulnerable financial behaviors in DeFi. The immaturity of the related technologies and the complicated economic environment have led to the difficulty of behavior detection in DeFi.

\textbf{(vii)} The security risks associated with cross-chain should be given more consideration. The cross-chain DeFi project aims to improve the transactional capabilities of the main chain. As shown in Table \ref{tab:blockchains}, however, we can identify the triangle balance problem. If the blockchain's scalability is increased, its security or decentralization may be compromised. Therefore, finding a balanced solution or solving the triangle balance problem is complex. However, it has significant implications for the development of DeFi.

\subsection{Future Directions}
In addition to cryptocurrencies, Ethereum and other systems in the 2.0 era of blockchain have a more comprehensive range of financial applications with the development of smart contracts. However, due to the combination of multiple protocol stacks and the development of blockchain technology, DeFi's security is also gaining attention. Therefore, we offer some possible future directions for enhancement.

\textbf{(i)} Oracle system that can adequately connect to the outside world is desperately needed. DeFi services are all about quantifying the laws of the natural world through technical means. However, when there is a large amount of data (asset) beyond what the quantified system can rate, this can cause the original quantified system to run out of steam. So DeFi desperately needs a sound oracle system to connect the objective external world to the DeFi system to achieve equal and reasonable coexistence between the data.

\textbf{(ii)} A secure DeFi protocol development experience guide is needed. As described in Section \ref{Security_Optimization}, many optimization options are available that improve security for developers and users. Moreover, with the growth of DeFi, more and more novices are getting into the DeFi protocol development work. And DeFi protocols' security optimization necessitates some prior knowledge. As a result, most novice developers cannot effectively use the optimization tools available. Therefore, urgent security guidance for developing DeFi protocols must promote the industry's steady growth.

\textbf{(iii)} Mature sidechain technology is worth developing. Financial projects involve many money transactions, and DeFi is no exception. Blockchain technology is crucial for DeFi services. When users initiate transactions using DeFi services, the blockchain underlay processes all the transactions before returning to the user interaction interface. So the transaction speed of blockchain technology is significantly correlated with the transaction speed of DeFi users. Moreover, without compromising security, the sidechain extension technology will better increase the system throughput of blockchain transactions to increase the number of transactions and transaction speed processed by the system.

\textbf{(iv)} The security of blockchain technology needs to be improved. As the underlying technology on which DeFi technology relies, blockchain technology needs more security research. As summarized in Section \ref{sec:ana_v}, many vulnerabilities still have caused real-world attacks. In Section \ref{sec:attack_events}, we can see that these attacks have caused severe damage, affecting people's confidence in DeFi technology. Although there have been many optimization techniques, the number of attacks against blockchain technology has not decreased. So it is worthwhile to continue researching security efforts for all layers of blockchain technology, especially the smart contract layer that triggered most of the vulnerabilities.

\textbf{(v)} Effective multi-layer vulnerability detection tool is still lacking. The detection method designed for a particular layer cannot detect higher-level information. For example, detection tools designed for the smart contract use the information in the smart contract, so they cannot detect the features at the application level. Most application-level attacks must combine multiple layers, so joint detection of vulnerabilities between multiple layers is worth investigating.

\textbf{(vi)} The dynamic supervision techniques for each layer need to be improved. There are many static analysis methods and detection techniques. However, these techniques cannot prevent the damage caused by the attack timely when the attack occurs. Efficient dynamic supervision technology could solve this dilemma. However, there is a lack of efficient supervision technology for various layers, including the data, network, and application layers.

\textbf{(vii)} DeFi application-layer inspection tools are insufficient. According to our survey, there are currently few excellent DeFi detection tools, such as the four shown in Table \ref{tab:DeFi_optimation}. However, automated detection of DeFi projects is crucial for project users and developers. The assets of users and developers are more secure for the project. In addition, we believe that combining application layer data with data from other layers improves application project security.

\textbf{(xvii)} Decentralized applications should be fully integrated with technical features. The current application design copies other existing frameworks, ensuring the application's usability but sacrificing security. For example, many DAPPs in DeFi simulate the real-world design of financial product data. However, they do not fully consider all the circumstances. A more significant number of assets than the existing assets suddenly entering the system will significantly shrink the existing proportion of assets. Thus, market participants cannot have the right to master the assets. Therefore, the DAPPs should take full advantage of the technical features instead of abandoning the design framework, thus leading to an imperfect combination.

\textbf{(xviii)} The study of indirect DeFi attacks deserves improvement. In Section \ref{sec:attack_events}, we compiled numerous actual DeFi attacks. Moreover, we discovered that they all correspond to vulnerabilities in Section \ref{sec:ana_v}. However, not all vulnerabilities in Section \ref{sec:ana_v} have disclosed attacks. It indicates that the damage caused by the vulnerabilities underlying DeFi applications does not immediately manifest at the application level. Vulnerabilities that target the underlying layer result in a state change at the application level, which leads to a series of attacks. 


\section{Conclusion}
\label{conclusion}

DeFi is a new type of platform based on blockchain technology that may increase the number of financial transactions while also efficiently enhancing the development of finance. This paper is the first systematic analysis of all levels of vulnerability, real-world attacks, and optimization schemes. Furthermore, based on our systematic analysis, we provide some DeFi challenges and future directions. First, we start with a systematic analysis of each layer, and a series of vulnerabilities are summarized. For each vulnerability, we investigate real-world attack cases and explore the vulnerabilities used in each case. We then summarize the studies on optimizing for these vulnerabilities at each layer. Finally, we summarize the dilemmas and security issues encountered. In terms of the future directions of optimization, we believe that comprehensive attack analysis and monitoring are critical to DeFi security.

\section*{Acknowledgement}
We thank the anonymous reviewers for their helpful comments. This research is partially supported by Early Career Research Starting Fund of Hainan University under Grant RZ2200001265.

\section*{Declaration of Interest}
The authors declare that they have no known competing financial interests or personal relationships that could have appeared to influence the work reported in this paper.

\printcredits

\bibliographystyle{model2-names.bst}

\bibliography{ref}

\end{document}